\documentclass[preprint,12pt]{elsarticle}

\usepackage{amssymb}
\usepackage{amsthm}
\usepackage{amsmath}
\usepackage{algorithm}
\usepackage{algpseudocode}
\usepackage{bm}
\usepackage[dvipsnames]{xcolor}
\usepackage{multirow}
\usepackage{soul}

\newtheorem*{remark}{Remark}
\newtheorem{proposition}{Proposition}
\usepackage{comment}
\usepackage{hyperref}
\begin{document}

\begin{frontmatter}



\title{Relax, estimate, and track: a simple battery state-of-charge and state-of-health estimation method}


\author{Shida Jiang$^a$, Junzhe Shi$^a$, Scott Moura$^a$}

\affiliation{organization={Department of Civil and Environmental Engineering},%Department and Organization
            addressline={University of California, Berkeley}, 
            city={Berkeley},
            postcode={94720}, 
            state={CA},
            country={USA}}

\begin{abstract}
Battery management is a critical component of ubiquitous battery-powered energy systems, in which battery state-of-charge (SOC) and state-of-health (SOH) estimations are of crucial importance. Conventional SOC and SOH estimation methods, especially model-based methods, often lack accurate modeling of the open circuit voltage (OCV), have relatively high computational complexity, and lack theoretical analysis. This study introduces a simple SOC and SOH estimation method that overcomes all these weaknesses. The key idea of the proposed method is to momentarily set the cell's current to zero for a few minutes during the charging, perform SOC and SOH estimation based on the measured data, and continue tracking the cell's SOC afterward. The method is based on rigorous theoretical analysis, requires no hyperparameter fine-tuning, and is hundreds of times faster than conventional model-based methods. The method is validated on six batteries charged at different C rates and temperatures, realizing fast and accurate estimations under various conditions, with a SOH root mean square error (RMSE) of around 3\% and a SOC RMSE of around 1.5\%. The data and codes are available at \url{https://berkeley.box.com/s/jz1w6po2iqzzfy7irxd9ok47ku3tr86j}.
\end{abstract}


\begin{highlights}
\item A simple SOC and SOH estimation method that requires no fine-tuning or initialization.
\item The method’s convergence and accuracy are guaranteed by detailed theoretical analysis.
\item Extremely low computation time ($<$ 1 ms) under various conditions.
\item A quantitative analysis of the impact of various error sources on the estimation.
\end{highlights}


\end{frontmatter}


\section{Introduction}
%background
Battery management is crucial for the operational efficiency, safety, reliability, and cost-effectiveness of ubiquitous battery-powered energy systems, such as electrified vehicles and smart grids with renewables \cite{start1}. Among different goals of battery management systems (BMS), battery state-of-charge (SOC) and state-of-health (SOH) estimation are of crucial importance for operating batteries \cite{start2}.

%definition of SOC & SOH
Battery SOC describes the actual energy level available at the battery and is defined as the ratio of the present available capacity to the present maximum capacity. On the other hand, battery SOH reflects the aging state of the battery and is defined as the ratio of the present maximum cell capacity (or present cell resistance) to its initial value \cite{SOH_definition}. Considering that the battery's internal resistance also changes as the battery is charged or discharged, we used the capacity version of the SOH definition in this paper. Namely, an 80\% SOH means that the cell's maximum capacity has decreased by 20\%. The main difference between the SOC and the SOH is that SOC indicates the instant status of the battery, while SOH indicates the long-term dynamic status of the battery \cite{difference}.

%rough review of SOH & SOC estimation methods
Many different methods have been proposed for SOC and SOH estimation. In general, these methods can be divided into three categories: direct measurement-based methods, data-driven methods, and physics-based methods. Direct measurement-based methods estimate the SOC and SOH through directly measurable features like voltage, current, and resistance. These methods generally have low computational complexity. However, they either have low accuracy (e.g., resistance method) or can be only used offline (e.g., ampere-hour counting method and impedance method) \cite{shida}. The only exceptions are the differential voltage method and its variant, the incremental capacity method, which yields relatively high accuracy SOC and SOH estimation results \cite{differntial1,differential2,differntial3}. The ``differential voltage'' refers to the derivative of the terminal voltage with respect to the capacity, and the ``incremental capacity'' refers to the derivative of capacity with respect to the terminal voltage. These two methods estimate the SOC and SOH by extracting related features from the differential voltage or incremental capacity curves. The limitations of these two methods are that they require precise voltage measurements, and they only work when the current is constant and is lower than a specific value \cite{yang2021review}.

On the other hand, data-driven methods estimate the battery SOC and SOH by training a black-box model with a large dataset \cite{smarter}. The input of the data-driven methods is usually health indicators derived from capacity, resistance, voltage, current, and temperature data \cite{AI}. The benefits of data-driven methods are that they do not need physical-based models and can have high accuracy \cite{shi2023robust}. The disadvantages are that they need high computational effort for training and are sensitive to the quantity and quality of training data \cite{SOH_definition}.

Meanwhile, physics-based methods estimate the SOC and SOH by first building a battery model to fit the raw data. Then, they use some model parameters to calculate the SOC and SOH indirectly. The battery models are sometimes formulated as battery equivalent circuit models (ECMs) or electrochemical models \cite{difference}, whose purpose is to predict the battery's voltage response to any input current. In model-based methods, SOC is usually estimated based on the state-space function and the SOC-OCV relationship. At the same time, SOH estimation can be done by using another filter or by calculating the derivative of SOC  \cite{ocv_SOC1,ocv_SOC2,ocv_SOC3}. In general, the accuracy of model-based methods depends on the model's accuracy. With the help of adaptive filtering algorithms such as a nonlinear Kalman filter (KF), physics-based methods can usually achieve good accuracy compared to other methods \cite{review_model}. The drawback is that the imperfection of the models often results in bias in the estimation, producing unexpected estimation errors. 

%online estimation challenge, model-based methods: OCV-SOC-SOH pipeline
However, although hundreds of different methods have been proposed for battery SOC and SOH estimation, we noticed that the existing methods have some common weaknesses. The first weakness is that, although the OCV-SOC relationship is widely used in SOC and SOH estimation, it also depends on temperature \cite{ocv_temp1,ocv_temp2,ocv_temp3} and SOH \cite{SOH_ocv, SOH_OCV2}. Yet, this influence (especially the effect of SOH) is ignored in most studies. Such a simplification can be problematic because the OCV-SOC function stored in the BMS will become increasingly inaccurate as the cell ages, making the SOC and SOH estimation results unreliable. Unfortunately, to the best of our knowledge, so far, only a few papers considered the effect of SOH when using the OCV-SOC relationship to estimate the SOC \cite{ocv1,ocv2,ocv3, SOHOCVnew1}. Worse still, in \cite{ocv1}, the authors only showed the relation between some parameters and SOH, yet no SOH estimation method was proposed or validated. In \cite{ocv2}, the primary focus was only SOC and OCV, and the SOH estimation result was not detailed. In \cite{ocv3}, although an SOH estimation method was proposed, the method is based on features from the OCV curve across a wide SOC range (from 10\% to 90\%), which are not fully accessible in many applications. In \cite{SOHOCVnew1}, the proposed method requires fine-tuning some hyperparameters and was only validated at 100\% and 96\% SOH, so its effectiveness still requires further investigation. In summary, while many papers used the OCV-SOC relationship to estimate the SOC and SOH, few have developed an effective way to integrate the effect of battery aging on OCV into their methods.

%battery pack challenge, the trade-off between accuracy and complexity
Another weakness of the existing battery SOC and SOH estimation literature is that little attention has been paid to reducing their computational complexity. In most academic studies, estimation accuracy is often the only metric used to measure how good or bad a method is. However, for state estimation in a battery pack, the computational complexity can also be an essential and practical factor to consider \cite{pack_review}. For example, in a Tesla Model S, there are 96 series-connected battery modules, so the workload and computational cost can be very high if we adopt a complex filter-based method for each module \cite{tesla}. To solve such a problem, we noticed that some studies proposed to use a simple method first to identify the ``weakest'' cell in the battery pack (i.e., the cell that has the lowest voltage or some other apparent characteristics that make it more likely to have the lowest SOC or SOH) and then use another complex method to estimate the state of those weakest cells \cite{tesla,pack1,pack2,preet}. While this idea can partially solve the problem, the method may not capture the weakest cell correctly, which could make the estimation too optimistic.

%lack of theoretical analysis
Furthermore, most SOC and SOH estimation methods so far only use experimental data for validation and do limited theoretical analysis. While empirical data may be enough to verify the method's effectiveness in one specific setting, such effectiveness can no longer be guaranteed once any setting (e.g., the sampling frequency of the voltmeter, the parameters of the ECM) changes. In the worst case, the algorithm may not even converge when the initial value is not accurate enough or when the measurement is not precise enough \cite{smarter}. For example, the differential voltage method often uses a filter to reduce voltage measurement noise. However, in most papers that use this method so far, the noise filtering algorithm is only proposed empirically \cite{differntial1,differential2,differntial3}, so it may not work well when the precision of the voltmeter is lower. As another example, the extended Kalman filter (EKF) and unscented Kalman filter (UKF) are often used for SOC and SOH estimation. However, since no theoretical analysis is done to guarantee their convergence in SOH estimation \cite{SOH_core}, it is unclear whether the method can always yield satisfactory results as the battery ages. 

%Main contributions
In this paper, we propose a novel and simple SOC and SOH estimation method that addresses all three weaknesses described above. The main contributions of this paper are:
\begin{itemize}
  \item We proposed a SOC and SOH estimation method that can be easily implemented online. The computational complexity of our method is extremely low ($<$ 1 ms), yet the estimation accuracy is rather high. Namely, the SOH root mean square error (RMSE) is about 3\%, and the SOC RMSE is about 1.5\%. 

  \item The proposed method has no hyperparameters and does not require initialization. The noise tolerance and the convergence of the method are guaranteed by detailed theoretical analysis.

  \item We conducted a very detailed error analysis to investigate the specific effect of all potential error sources. The conclusions can also be applied to other ECM-based SOC and SOH estimation methods.
\end{itemize}

The paper is organized as follows. Section 2 introduces the definitions of OCV, SOC, and SOH used in the paper. Section 3 presents the details of our SOC and SOH estimation method. Section 4 provides the theoretical basis of our method. Section 5 presents the experimental results and compares our method against UKF. A detailed analysis of the estimation error is also given in Section 5. Finally, in Section 6, we discuss the conclusions drawn from this study.

\section{SOC, SOH, and OCV}
\subsection{Definition of SOC and SOH}
The battery SOC is defined as:
\begin{equation}
\label{SOC}
    SOC=\frac{Q_{r}}{Q} \times 100\%
\end{equation}
where $Q$ is the present maximum capacity of the battery, and $Q_{r}$ is the remaining capacity of the battery.

The derivative form of (\ref{SOC}), which is often used for SOC estimation, is:
\begin{equation}
\label{dSOC}
    \frac{dSOC}{dt}=\frac{I}{Q}
\end{equation}

On the other hand, the battery SOH is defined as:
\begin{equation}
\label{SOH}
    SOH=\frac{Q}{Q_0} \times 100\%
\end{equation}
where $Q_0$ is the maximum capacity of a new cell.

It is worth mentioning that a cell's capacity varies at different C rates. For consistency, all the ``capacity'' values above refer to the charge capacity (calculated by Coulomb counting) when the C rate is 0.1 C.

\subsection{OCV model}
\label{OCV_section}
In this study, the OCV curves were fitted by a ninth-order polynomial function due to its simplicity and low RMS error \cite{OCVmodel_compare}.
\begin{equation}
\label{OCV}
    OCV=f(SOC,SOH,T)=\sum_{i=0}^{9}a_i(SOH,T) \cdot SOC^i
\end{equation}
where $a_i,i=0,1,...,9$ are coefficients related to SOH and temperature (denoted by $T$ in the equation). These coefficients can be determined by fitting the OCV curve of the battery. 

\begin{figure}[htbp]
\centering
\includegraphics[width=10cm]{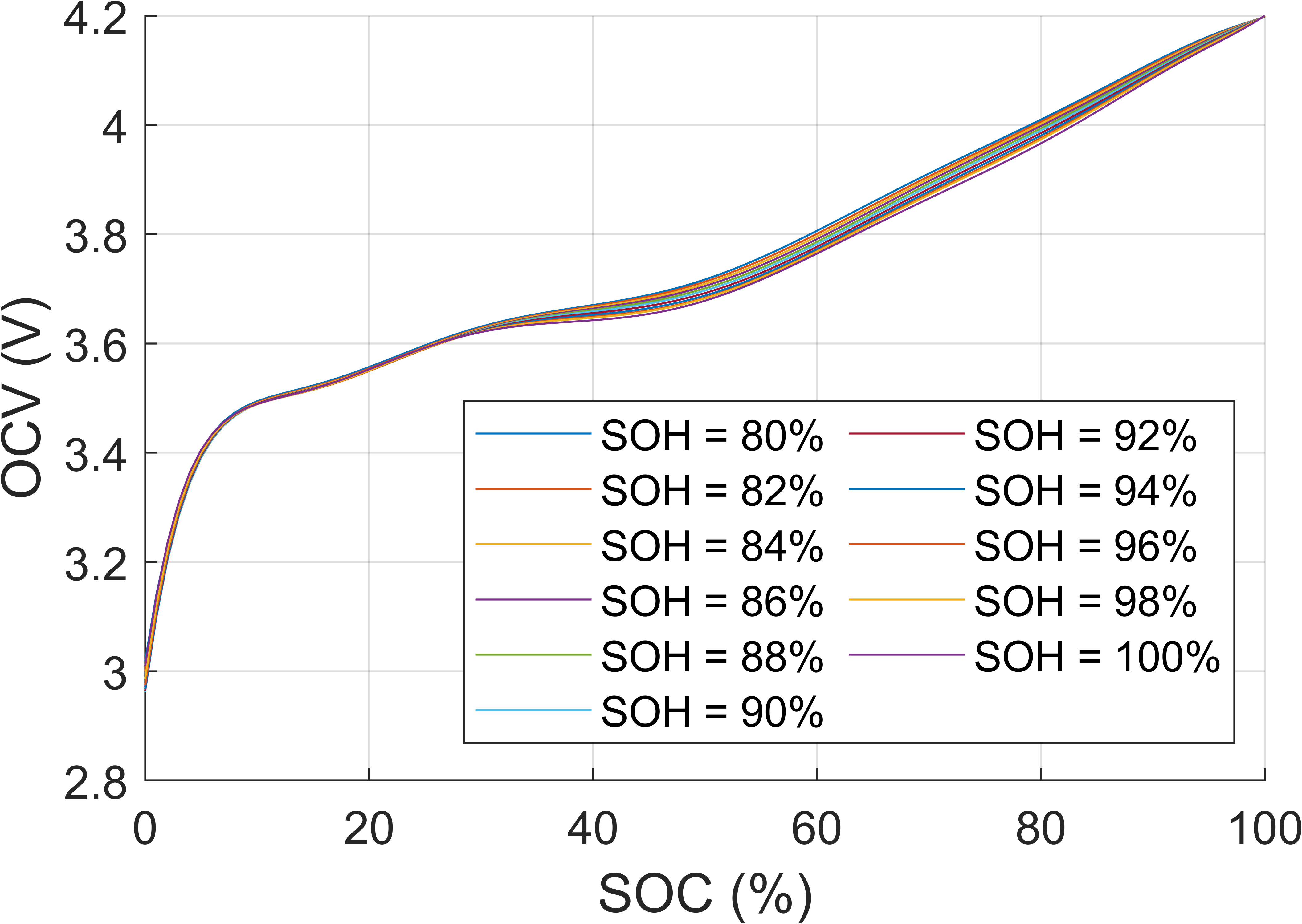}
\caption{The fitted OCV curve at 25$^\circ$C}
\label{OCVcurve}
\end{figure}

There are generally two ways to acquire the OCV curve in experiments. The first is the slow-current OCV test, and the second is the incremental OCV test. In a slow-current OCV test, the cell is fully discharged and then fully charged at a constant current that is lower or equal to 0.1 C. On the other hand, in an incremental OCV test, the cell is usually charged and discharged with a higher C rate (e.g., 0.5 C). Whenever the SOC rises or drops by a certain percentage (e.g., 10\%), the cell is open-circuited for some time (e.g., two hours) \cite{OCVtest}. Compared with the low-current OCV test, the incremental OCV test has been shown to describe the battery behavior better \cite{OCVtest} and make model-based SOC estimation more accurate \cite{OCVtest2}. As a result, in this study, incremental OCV tests are used to acquire the OCV-SOC curve. The OCV curve acquired from fitting the experimental data at 25$^\circ$C is presented in Figure \ref{OCVcurve}.

Note that the OCV curve in Figure \ref{OCVcurve} is only used in the theoretical analysis in Section \ref{converge} and never used in the experimental validation in Section \ref{section_val}. In the experimental validation, when the estimation is validated on a particular cell, that cell's OCV data will be excluded when fitting the OCV curve. The purpose is to separate the data used for fitting the OCV curve and the data used for experimental validation.
\section{Methodology}
\label{parameter_section}
This section will discuss the details of the SOC and SOH estimation method. Since the SOH changes much more slowly than SOC, it can be considered constant during a single cycle. With this simplification and the definition of SOH in (\ref{SOH}), when the temperature is constant, (\ref{dSOC}) can be rewritten as follows:
\begin{equation}\label{SOH2}
    SOH=\frac{I}{Q_0}\frac{dt}{dSOC}=\frac{I}{Q_0}\frac{\partial OCV}{\partial SOC}\frac{dt}{dOCV}
\end{equation}
Equations (\ref{OCV}) and (\ref{SOH2}) represent two independent equations and two unknowns: SOC and SOH. Given these two equations, we can solve the values of SOC and SOH iteratively. Namely, as shown in (\ref{iteration1}), the SOH is initially set to 100\%, and (\ref{OCV}) is used to solve the value of SOC. Next, the value of SOC is substituted into (\ref{SOH2}) to update the SOH. Afterward, the two processes described above are repeated until both estimations converge. An analysis of the convergence properties is detailed in Section \ref{converge}.
\begin{equation}\label{iteration1}
    \begin{cases}SOH_0=1\\SOC_{k}=f^{-1}(OCV,SOH_k,T)\\SOH_{k+1}=\frac{I}{Q_0} \frac{\partial OCV}{\partial SOC}\frac{dt}{dOCV}\end{cases} 
\end{equation}
To compute (\ref{iteration1}), we need values for $OCV$ and $dOCV / dt$ at a certain $SOC$.
% To ensure that $SOC_k$ and $SOH_k$ converge to their true value using the iteration in (\ref{iteration1}), the following three conditions need to be satisfied. First, we know the value of OCV at a certain SOC; second, we know the value of dOCV/dt at the same SOC; and third, the iterative algorithm can converge. In this section, we mainly examine the first two conditions, that is, estimating $OCV$ and $dOCV/dt$. The third condition is that the mapping from $SOH_k$ to $SOH_{k+1}$ in (\ref{iteration1}) is (locally) a contraction mapping, as detailed in Section \ref{converge}.

\begin{figure}[htbp]
\centering
\includegraphics[scale=0.4]{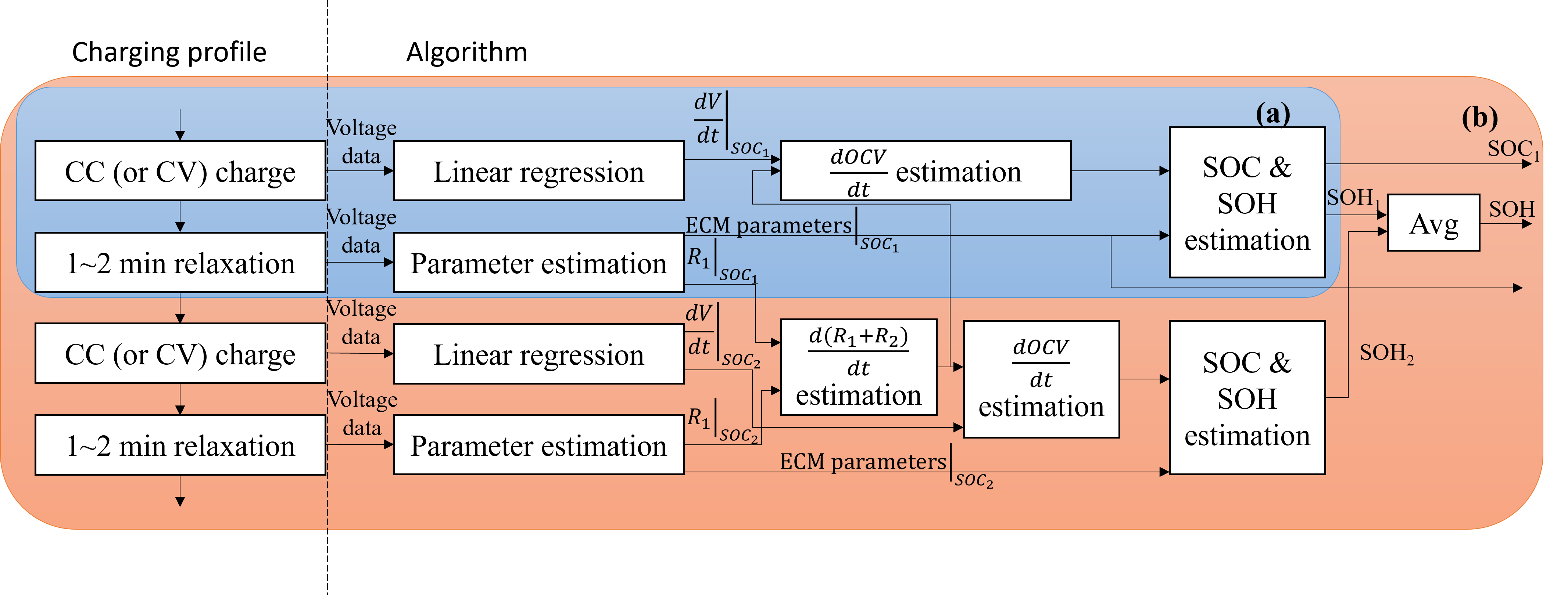}
\caption{A flow chart of the proposed SOC and SOH estimation method: (a) without $dR$ compensation (b) with $dR$ compensation}
\label{algorithm}
\end{figure}

While the basic idea of the iterative estimation method is straightforward, estimating $OCV$ and $dOCV/dt$ at the same SOC is not an easy task. From the definition, $OCV$ is only directly measurable when the current is zero. However, when the current is zero, $dOCV/dt$ is also zero, making (\ref{SOH2}) meaningless. Therefore, we propose to implement the estimation method during a carefully designed charging profile, which is a constant-current-constant-voltage (CCCV) charging profile with short relaxations in the middle. Specifically, as shown in Figure \ref{algorithm}(a), the current and voltage data during the CCCV charging are used to estimate $dOCV/dt$ right before entering the relaxation period; the data during the relaxation period are used to estimate the $OCV$. Because the SOC is constant during the relaxation period, these two estimations correspond to the same SOC, satisfying the first two prerequisites we mentioned previously. Note that the algorithm shown in Figure \ref{algorithm} has two variants. When the current during the CCCV charging is not high ($\leq0.5$ C), the first variant is enough for accurate SOC and SOH estimation. Meanwhile, when the current is higher than 0.5 C, the second variant, which performs the parameter estimation twice, is more suitable. The reason for introducing the second variant and the detailed process of estimating $OCV$ and $dOCV/dt$ are discussed in depth in the following two subsections.

\subsection{Estimating the open-circuit voltage}
\begin{figure}[htbp]
\centering
\includegraphics[width=8cm]{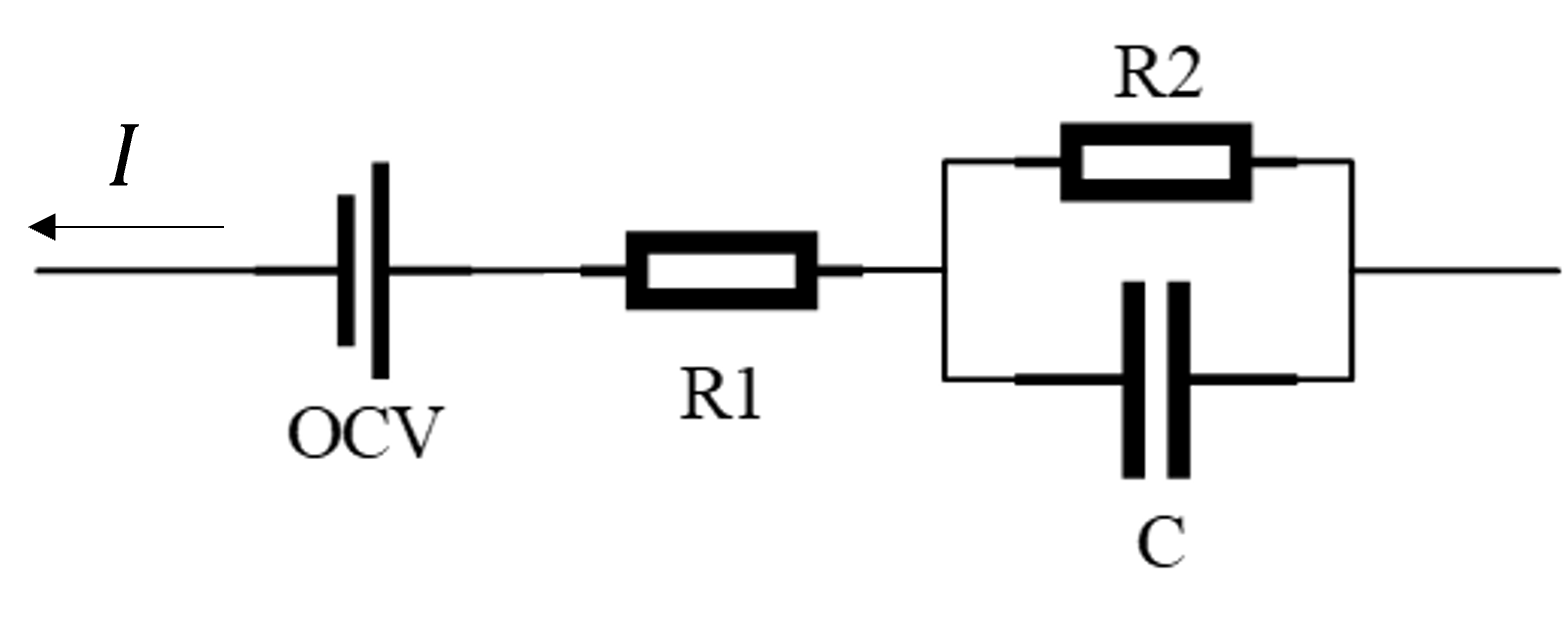}
\caption{Battery equivalent circuit model}
\label{ECM}
\end{figure}

The $OCV$ of a battery is not directly measurable unless it is idle for hours. Therefore, an ECM is usually required to estimate this parameter online. The ECM used in this paper is presented in Figure \ref{ECM}. Besides the OCV element, the ECM consists of a resistor and an RC pair (a resistor and a capacitor connected in parallel).

If we select SOC and capacitor voltage ($U_c$) as the states, current ($I$) as the input, and terminal voltage ($U_{ter}$) as the output, the state-space model can be written as
\begin{equation}
\label{state_space0}
    \begin{bmatrix} \dot{SOC}  \\ {\dot{U}_c} \end{bmatrix} = \begin{bmatrix}0 & 0 \\0 & \frac{-1}{R_2C} \end{bmatrix} \begin{bmatrix}SOC  \\U_c \end{bmatrix} +\begin{bmatrix}\frac{1}{Q}   \\\frac{1}{C} \end{bmatrix} I
\end{equation}
\begin{equation}
\label{output}
    U_{ter}=OCV(SOC,SOH,T)+U_c+R_1I
\end{equation}
where $R_1$, $R_2$, and $C$ are the ECM parameters shown in Figure \ref{ECM}.

 Suppose that at time $t_0$, the current suddenly drops to zero, then the output function becomes
\begin{equation}
\label{exp}
    U_{ter}(t_0+\Delta t)=OCV(SOC,SOH,T)+U_{c}(t_0)e^{-\frac{\Delta t}{RC}}, \Delta t>0
\end{equation}
If the current before $t_0$ was constant (denoted as $I_0$) and the capacitor had already reached steady-state, then 
\begin{equation}
\label{Uc}
    U_{c}(t_0)=I_0R_2
\end{equation}
As for the parameter $R_1$, its value can be estimated by using the equation below
\begin{equation}
\label{R1}
    R_1=-\Delta U_{ter}/I_0
\end{equation}
where $\Delta U_{ter}$ is the sudden change in terminal voltage after the current becomes zero.

Given (\ref{exp}), (\ref{Uc}), and (\ref{R1}), all the parameters in the ECM can be estimated based on the measurement data. Namely, $R_1$ can be directly calculated by (\ref{R1}), and the other three parameters can be identified using regression analysis or parameter estimation techniques. However, these algorithms are relatively computationally intensive and unsuitable for cell-level or module-level estimation in a large battery pack with thousands of cells and around a hundred modules. Consequently, a more straightforward parameter estimation method is used in this paper to estimate the other three ECM parameters, which is to determine the three parameters by using three data points. Namely, given three pairs of time and terminal voltage data $(x_1, y_1)$, $(x_2, y_2)$, and $(x_3, y_3)$, the parameters can be determined by solving the following set of equations
\begin{equation}
\label{temp3}
\begin{cases}y_1=OCV+ I_0R_2e^{-\frac{x_1}{R_2C}}\\y_2=OCV+ I_0R_2e^{-\frac{x_2}{R_2C}}\\y_3=OCV+ I_0R_2e^{-\frac{x_3}{R_2C}}\end{cases} 
\end{equation}
where $x_1,x_2,x_3$ are time durations after the relaxation starts. 

To improve the noise tolerance of the method, the selection of the three points must be optimized. Otherwise, a small noise in the measurement can make the parameter inaccurate. For this purpose, the general rule of thumb is to select a small $x_1$, a big $x_3$, and select $x_2=\frac{x_1+x_3}{2}$. The theoretical analysis behind this rule is explained in Section \ref{3points}. Following this rule, if we denote $x_3-x_2=x_2-x_1=x_d$, then (\ref{temp3}) can be transformed into:
\begin{equation}
\label{final3}
\begin{cases}
R_2=\frac{(y_1-y_2)e^{x_d/\tau}}{I_0(e^{x_d/\tau}-1)}\\
C=\frac{\tau}{R_2}\\
OCV=y_1-\frac{(y_1-y_2)e^{x_d/\tau}}{e^{x_d/\tau}-1}
\end{cases} 
\end{equation}
where $\tau=\frac{x_d}{\ln\frac{y_1-y_2}{y_2-y_3}}$ and $\tau$ can be interpreted as the time constant of the RC pair.

Based on (\ref{final3}), the values of the three parameters can be directly calculated once the three data points are acquired. However, even if the three points have been optimized, estimating three parameters using three data points is still inevitably vulnerable to measurement noise. Therefore, to further improve the noise tolerance, it is necessary to filter out the noise before doing parameter estimation. The algorithm we chose here is sorting, which sorts the voltage measurements so that they become monotonous during the relaxation. We chose this algorithm instead of other common filtering algorithms like averaging for two reasons. First, this algorithm is straightforward and adds minimal computational complexity. Second, we can theoretically prove that this filtering algorithm can always reduce the variance of the noise. In contrast, other algorithms only work when the variance of the noise is in a specific range. Namely, we have the following proposition.
\begin{proposition}
\label{swap}
Suppose that $x_1>x_2$, yet the estimation of $x_1$ (denoted as $\hat{x}_1$), is smaller than the estimation of $x_2$ (denoted as $\hat{x}_2$), i.e. $\hat{x}_1 < \hat{x}_2$. Define the mean-square error (MSE) of the two estimations as $0.5[(x_1-\hat{x}_1)^2+(x_2-\hat{x}_2)^2]$. In this case, swapping the two estimations (i.e., use $\hat{x}_1$ as the estimation of $x_2$ and $\hat{x}_2$ as the estimation of $x_1$) will make the MSE smaller. 
\end{proposition}
\begin{proof}
After swapping, the change in MSE is
\begin{align*}
    &0.5[(x_1-\hat{x}_2)^2+(x_2-\hat{x}_1)^2]-0.5[(x_1-\hat{x}_1)^2+(x_2-\hat{x}_2)^2]\\
    =&0.5(-2\hat{x}_2x_1-2\hat{x}_1x_2+2\hat{x}_1x_1+2\hat{x}_2x_2)\\  
    =&(\hat{x}_1-\hat{x}_2)(x_1-x_2)<0
\end{align*}
Meaning that the MSE becomes smaller after the swapping.
\end{proof}
\begin{remark}
Proposition \ref{swap} means that swapping two measurements that are not in order can reduce measurement noise. In other words, the noise can be minimized by sorting all measured data, irrespective of timestamp.
\end{remark}
\subsection{Estimating the derivative of open-circuit-voltage}\label{method_section}
Once the OCV during the relaxation is identified, the only missing piece in the SOC and SOH estimation method represented by (\ref{iteration1}) is estimating $\frac{dOCV}{dt}$. When the capacitor in the ECM reaches a steady state, the terminal voltage can be calculated by:
\begin{equation}\label{Uter}
    U_{ter}=OCV+I(R_1+R_2)
\end{equation}
If we further know that the current is constant and is equal to $I_0$, the derivative of (\ref{Uter}) can be calculated as:
\begin{equation}
\label{dV}
    \frac{dU_{ter}}{dt}=\frac{dOCV}{dt}+I_0\frac{dSOC}{dt}(\frac{dR_1}{dSOC}+\frac{dR_2}{dSOC})
\end{equation}
Both $\frac{dOCV}{dt}$ and $\frac{dSOC}{dt}$ are proportional to $I_0$ because $\frac{d OCV}{dt} = \frac{d OCV}{d SOC} \cdot \frac{I_0}{Q}$ using chain rule. Therefore, when the current is small, we can neglect the second term on the right-hand side of (\ref{dV}) since it is proportional to $I_0^2$, and (\ref{dV}) becomes:
\begin{equation}
\label{approx}
    \frac{dU_{ter}}{dt} \approx \frac{dOCV}{dt}
\end{equation}

Equation (\ref{approx}) points out that $\frac{dU_{ter}}{dt}$, which can be calculated through numerical differentiation, approximately equals to $\frac{dOCV}{dt}$ when the current is small and constant. One thing to note is that $\frac{dU_{ter}}{dt}$ is related to SOC. So, to ensure the SOC is the same as the SOC in parameter estimation, the voltage data right before the relaxation is used to do the regression analysis and determine $\frac{dU_{ter}}{dt}$ and $\frac{d OCV}{dt}$. 

Besides constant-current(CC) charging, our algorithm can also be implemented during constant-voltage (CV) charging. In this case, noticing that $\frac{dU_{ter}}{dt}=0$, the derivative of (\ref{Uter}) is
\begin{equation}
\label{dI}
    \frac{dOCV}{dt}+\frac{dI}{dt}(R_1+R_2)+I\frac{dSOC}{dt}(\frac{dR_1}{dSOC}+\frac{dR_2}{dSOC})= 0
\end{equation}
Since the current during the CV charging is usually small, we can neglect the third term on the left-hand side of (\ref{dI}) since it is proportional to $I^2$, and (\ref{dI}) becomes:
\begin{equation}
\label{dI2}
    \frac{dOCV}{dt}+\frac{dI}{dt}(R_1+R_2)\approx 0
\end{equation}
Since $R_1$ and $R_2$ can be estimated by using (\ref{R1}) and (\ref{final3}), the value of 
$\frac{dOCV}{dt}$ during the constant-voltage charge can be easily estimated by using (\ref{dI2}). When $dOCV/dt$ is estimated by either (\ref{approx}) or (\ref{dI2}), the entire SOC and SOH estimation method is shown in Figure \ref{algorithm}(a).

However, when the current is high (above 0.5 C), the approximation in (\ref{approx}) would be no longer accurate since $I\frac{d(R_1+R_2)}{dt}$ will not be negligible. To address this problem, we propose another variant of the SOC and SOH estimation method, which performs the parameter estimation a second time after the cell charges for a short while. While this will double the total relaxation time and calculation complexity, it can bring two benefits. First and foremost, by comparing the results from the first and second parameter estimation, we can know how much $(R1+R2)$ has changed during that short period and calculate $d(R1+R2)/dt$, which can help us better estimate $dOCV/dt$ through (\ref{dV}). The second benefit is that the estimation result can be more accurate if the estimation is done twice and averaged. A flow chart of the second variant of the proposed SOC and SOH estimation method is presented in Figure \ref{algorithm}(b). For both variants, the SOC in the output is the cell's SOC in the first relaxation. Since SOC changes fast, another algorithm (such as coulomb counting) is usually needed to keep tracking the SOC afterward. We call this final step "tracking". Note that the SOC tracking method can be very simple because the SOC and SOH of each cell during the relaxation have already been estimated. An example of such a SOC tracking method can be found in \ref{section_track}. Generally, the proposed method only requires simple matrix operations and has no hyperparameters. The only additional requirement is some extra relaxations during CC (or CV) charging, which only delays the charging by no more than four minutes.

\section{Theoretical analysis}
\subsection{Sensitivity analysis}
\label{3points}
In section \ref{parameter_section}, when the three measurements $y_1,y_2,y_3$ are noisy, the estimation of the three ECM parameters $R_2,C,OCV$ can also be inaccurate. When such noise is small, the relationship between the measurement error and the parameter estimation error can be found by taking the derivative of (\ref{temp3}) and re-organizing into,
\scriptsize
\begin{equation} 
    \begin{bmatrix} \frac{dR_2}{R_2}  \\ \frac{dC}{C}\\ \frac{dOCV}{OCV} \end{bmatrix} = \begin{bmatrix}
    \frac{x_3e^{bx_3}-x_2e^{bx_2}}{a(A_1+A_2-A_3)} & 
    \frac{x_1e^{bx_1}-x_3e^{bx_3}}{a(A_1+A_2-A_3)} & 
    \frac{x_2e^{bx_2}-x_1e^{bx_1}}{a(A_1+A_2-A_3)} \\ 
    \frac{(bx_2-1)e^{bx_2}-(bx_3-1)e^{bx_3}}{ab(A_1+A_2-A_3)} & 
    \frac{(bx_3-1)e^{bx_3}-(bx_1-1)e^{bx_1}}{ab(A_1+A_2-A_3)} & 
    \frac{(bx_1-1)e^{bx_1}-(bx_2-1)e^{bx_2}}{ab(A_1+A_2-A_3)}\\ 
    \frac{1}{OCV}\frac{A_2}{A_1+A_2-A_3} &
    \frac{1}{OCV}\frac{-A_3}{A_1+A_2-A_3} & 
    \frac{1}{OCV}\frac{A_1}{A_1+A_2-A_3} 
    \end{bmatrix} 
    \begin{bmatrix} dy_1  \\ dy_2 \\dy_3 \end{bmatrix}
\end{equation}
\normalsize
where $a=R_2I_0$, $b=-\frac{1}{R_2C}<0$, $A_1=(x_2-x_1)e^{\frac{x_3-x_2}{R_2C}}$, $A_2=(x_3-x_2)e^{\frac{x_1-x_2}{R_2C}}$, and $A_3=x_3-x_1$.

Since the three measurements are done using the same voltmeter, we can assume that the variance of the three measurements are the same and are all equal to $\sigma_{y}^2$. In this case, the variance of the estimations of the three parameters are
\begin{align}
    &\sigma_{R_2}^2=\frac{(x_3e^{bx_3}-x_2e^{bx_2})^2+(x_1e^{bx_1}-x_3e^{bx_3})^2+(x_2e^{bx_2}-x_1e^{bx_1})^2}{a^2(A_1+A_2-A_3)^2}R_2^2\sigma_{y}^2\\
    &\sigma_{C}^2=\frac{(b_2e^{bx_2}-b_3e^{bx_3})^2+(b_3e^{bx_3}-b_1e^{bx_1})^2+(b_1e^{bx_1}-b_2e^{bx_2})^2}{a^2b^2(A_1+A_2-A_3)^2}C^2\sigma_{y}^2\\
    &\sigma_{OCV}^2=\frac{A_1^2+A_2^2+A_3^2}{(A_1+A_2-A_3)^2}\sigma_{y}^2 
    \label{std_OCV}
\end{align}
where $b_i=bx_i-1,i=1,2,3$.

Since the final purpose of parameter estimation is to estimate the SOC and SOH, and only OCV is directly related to the SOC and SOH, it is only necessary to minimize the covariance of OCV estimation. In other words, if we define $f(x_1,x_2,x_3)=\sigma_{OCV}^2/\sigma_{y}^2$, the objective function of the optimization problem can be formulated as $\displaystyle \min_{x_1,x_2,x_3} f$. The following propositions are introduced to solve this optimization problem.
\begin{proposition}
\label{infx}
 The following three limits hold:
 \begin{enumerate}

     \item As $x_3-x_1 \to 0$, then $f\to \infty$
     \item As $x_3-x_2 \to \infty$, then $f \to 1$
     \item As $x_2-x_1 \to \infty$, then $f \to \frac{1+\exp\left(\frac{2(x_3-x_2)}{R_2C}\right)}{(\exp\left(\frac{x_3-x_2}{R_2C}\right)-1)^2}$.
 \end{enumerate}
\end{proposition}
\begin{proof}
When $x_3-x_1 \to 0$, both $x_2-x_1$ and $x_3-x_2$ also $\to 0$. So,
\begin{equation}
A_1+A_2-A_3=o(x_d)
\end{equation}
where $x_d=\max(x_2-x_1,x_3-x_2)$. While
\begin{equation}
    A_1^2+A_2^2+A_3^2=O(x_d^2)
\end{equation}
So,
\begin{equation}
    f=\frac{A_1^2+A_2^2+A_3^2}{(A_1+A_2-A_3)^2}=\frac{O(x_d^2)}{o(x_d^2)} \to \infty
\end{equation}
On the other hand, when $x_3-x_2 \to \infty$, $A_1\gg A_3\gg A_2$, so
\begin{equation}
    f=\frac{A_1^2+A_2^2+A_3^2}{(A_1+A_2-A_3)^2}\approx \frac{A_1^2}{A_1^2}=1 
\end{equation}
Meanwhile, when $x_2-x_1 \to \infty$, $A_1=e^{\frac{x_3-x_2}{R_2C}}A_3>>A_2$, so
\begin{equation}
    f=\frac{A_1^2+A_2^2+A_3^2}{(A_1+A_2-A_3)^2}\approx \frac{1+e^{\frac{2(x_3-x_2)}{R_2C}}}{(e^{\frac{x_3-x_2}{R_2C}}-1)^2}
\end{equation}
\end{proof}
Proposition \ref{infx} suggests that increasing the time interval between the second and third points can improve the algorithm's noise tolerance. However, when the time interval is larger, the required relaxation time for the cell will be longer, which would cause more delay in the charging process. As a result, it is necessary to optimize the position of $x_2$ when the total time interval is constrained. To address this issue, we have the following proposition.

\begin{figure}[ht]
\centering
\includegraphics[width=10cm]{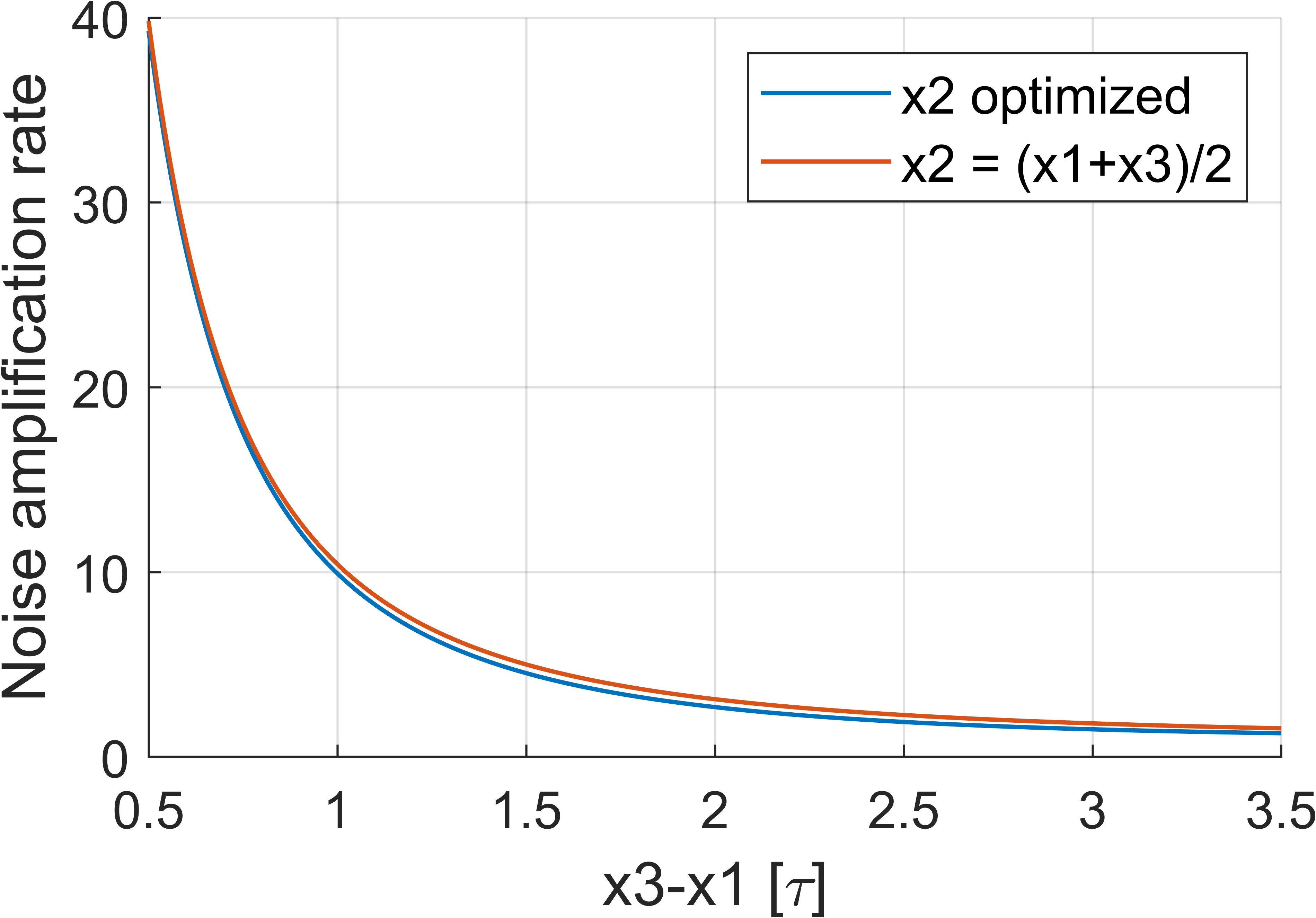}
\caption{The noise amplification rate at different $x_3$ (when $x_1=0$), $\tau=R_2C$}
\label{x3_tradeoff}
\end{figure}

\begin{proposition}
\label{x2}
When $x_1$ and $x_3$ are fixed, and $x_3-x_1 \ll R_2C$, the minimum of $\sigma_{OCV}^2/\sigma_{y}^2$ is obtained when $x_2=\frac{x_1+x_3}{2}$.
\end{proposition}
\begin{proof}
When $x_3-x_1 \ll R_2C$, $x_3-x_2$ and $x_2-x_1$ are also $\ll R_2C$, so
\begin{align}
    \frac{\partial A_1}{\partial x_2}&=1-2kx_2+kx_1+kx_3+o(x_3-x_1)\label{refequ1}\\
    \frac{\partial A_2}{\partial x_2}&=-1+2kx_2-kx_1-kx_3+o(x_3-x_1)\label{refequ2}\\
    \frac{\partial A_3}{\partial x_2}&=0\label{refequ3}
\end{align}
As a result, $\frac{\partial A_1}{\partial x_2}+\frac{\partial A_2}{\partial x_2}=o(x_3-x_1)$. When $f$ is minimized, $\frac{\partial f}{\partial x_2}=0$, so
\begin{equation} \label{targetequ}
(2A_1 \frac{\partial A_1}{\partial x_2}+2A_2\frac{\partial A_1}{\partial x_2})(A_1+A_2-A_3)-2(A_1^2+A_2^2+A_3^2)(\frac{\partial A_1}{\partial x_2}+\frac{\partial A_2}{\partial x_2})=0
\end{equation}
Substituting (\ref{refequ1}) and (\ref{refequ2}) into (\ref{targetequ}), we have:
\begin{equation}\label{finalequ}
    \frac{\partial A_1}{\partial x_2}(A_1-A_2)(A_1+A_2-A_3)=o((x_3-x_1)^3)
\end{equation}
If $\frac{2x_2}{x_1+x_3}\neq 1$, then $\frac{\partial A_1}{\partial x_2}=O(1)$, $A_1-A_2=O(x_3-x_1)$, $A_1+A_2-A_3=O(x_3-x_1)$, and (\ref{finalequ}) cannot be satisfied. As a result, when $\frac{\partial f}{\partial x_2}=0$, $\frac{2x_2}{x_1+x_3}=1$ must be satisfied. Noticing that this is the only point where $\frac{\partial f}{\partial x_2}=0$, $f$ must either reach maximum or minimum. Since when $x_2=x_1$ or when $x_2=x_3$, $f=+\infty$, we can conclude that $f$ is minimized at $x_2=\frac{x_1+x_3}{2}$ when $x_3-x_1$ is small.
\end{proof}
While Proposition \ref{x2} suggests that the optimal selection of $x_2$ is $\frac{x_1+x_3}{2}$, such a conclusion is based on the assumption that $x_3$ is small. To study if choosing $x_2=\frac{x_1+x_3}{2}$ is still reasonable when $x_3$ is large, in Figure \ref{x3_tradeoff}, a comparison of the noise amplification rate is made between the two $x_2$ selection schemes. The first scheme is to optimize $x_2$ so that the noise amplification rate is the lowest, while the second scheme is to select $x_2$ as $\frac{x_1+x_3}{2}$. It can be seen that the two curves are very similar, meaning that selecting $x_2$ as $\frac{x_1+x_3}{2}$ produces a near optimal result. 

In fact, selecting $x_2=\frac{x_1+x_3}{2}$ also has another benefit: the function set (\ref{temp3}) can have a closed-form solution and can be directly calculated by using (\ref{final3}). Meanwhile, if $x_2$ is chosen arbitrarily, the parameter needs to be calculated using algorithms like gradient descent, which damages the algorithm's simplicity. Considering all the above, $x_2$ is selected as $\frac{x_1+x_3}{2}$ in this paper. 

Finally, it is necessary to study how $f$ changes as $x_1$ and $x_3$ change. Regarding this, we provide the following proposition.
\begin{proposition}
\label{x3}
When $x_2=\frac{x_1+x_3}{2}$, $f$ decreases as $x_3$ increases or $x_1$ decreases. 
\end{proposition}
\begin{proof}
If we define $x_d=x_3-x_2=x_2-x_1$, $f$ would only be related to $x_d$, and the proposition we are trying to prove becomes equivalent to $\frac{df}{dx_d}<0$. And we have
\begin{equation}
    A_1=x_de^{\frac{x_d}{R_2C}}, 
    A_2=x_de^{-\frac{x_d}{R_2C}},
    A_3=2x_d
\end{equation}
Therefore, if we define $m=e^{\frac{x_d}{R_2C}}>1$, we have
\begin{align*}
    f&=\frac{A_1^2+A_2^2+A_3^2}{(A_1+A_2-A_3)^2}\\
    &=\frac{m^2+m^{-2}+4}{(m+m^{-1}-2)^2}\\
    &=\frac{m^4+4m^2+1}{(m-1)^4}
\end{align*}
Consequently,
\begin{equation}
    \frac{df}{dm}=\frac{-8m^3-16m-4}{(m-1)^5}<0
\end{equation}
Since $\frac{dm}{dx_d}>0$, it can be concluded that $\frac{df}{dx_d}<0$, which means that a larger $x_3$ or a smaller $x_1$ can make $f$ smaller.
\end{proof}

\subsection{Convergence analysis}
\label{converge}
In (\ref{iteration1}), an iterative method was proposed for SOC and SOH estimation. 
While the method can theoretically be applied at any time during the charging process, an additional concern is its convergence. Since the OCV curve is a nonlinear function of SOC and SOH (for a constant temperature), it is unclear if the process converges. Additionally, if it converges, the convergence may be faster in a specific SOC range. Therefore, we seek to analyze the convergence of the method and identify the best SOC range to use it in terms of convergence speed. 

Suppose that the true SOC and SOH are respectively $SOC_{true}$ and $SOH_{true}$, according to (\ref{iteration1}),
\begin{equation}\label{iteration_formular}
    SOH_{true}=\frac{I}{Q_0} \frac{dt}{dOCV}\left. \frac{\partial OCV}{\partial SOC}\right |_{SOC_{true}, SOH_{true}}
\end{equation}
The iteration process in (\ref{iteration1}) can therefore be written as
\begin{equation}\label{iteration_formular2}
    SOH_{k+1}=g(SOH_{k})=\frac{SOH_{true}\left. \frac{\partial OCV}{\partial SOC}\right |_{SOC_{k}, SOH_{k}}} {\left. \frac{\partial OCV}{\partial SOC}\right |_{SOC_{true}, SOH_{true}}}
\end{equation}
For simplicity, we define $\alpha=SOH_{true}/\left. \frac{\partial OCV}{\partial SOC}\right |_{SOC_{true}, SOH_{true}}$, and (\ref{iteration_formular2}) can be rewritten as:
\begin{equation}
    SOH_{k+1}=g(SOH_{k})=\alpha\left. \frac{\partial OCV}{\partial SOC}\right |_{SOC_{k}, SOH_{k}}
\end{equation}
We define the iteration to be locally convergent at $A=(SOC_{true}, SOH_{true})$ when there exists a neighborhood of $A$, and the iterative method will converge to $A$ when the initial point is in the neighborhood.  The local convergence of the method can be assessed by analyzing the derivative of $g$ with respect to its argument, which is defined in (\ref{L}).
\footnotesize
\begin{equation}\label{L}
    L=\left. \frac{dg}{dSOH} \right |_{SOC_{true}, SOH_{true}}=\alpha\big(\left. \frac{\partial^2 OCV}{\partial SOC^2}\frac{\partial SOC}{\partial SOH}+\frac{\partial^2 OCV}{\partial SOC \partial SOH}\big)\right |_{SOC_{true}, SOH_{true}}
\end{equation}
\normalsize
which assumes that $SOC_k$ and $\frac{dOCV}{dSOC}$ are differentiable w.r.t. $SOH_k$, and where
\begin{equation}
    \left. \frac{\partial SOC}{\partial SOH}\right |_{SOH_{true},SOC_{true}}=\left.\frac{\partial f^{-1}(OCV,SOH,T)}{\partial SOH}\right |_{SOH_{true},SOC_{true}}
\end{equation}

\begin{figure}[htbp]
\centering
\includegraphics[width=10cm]{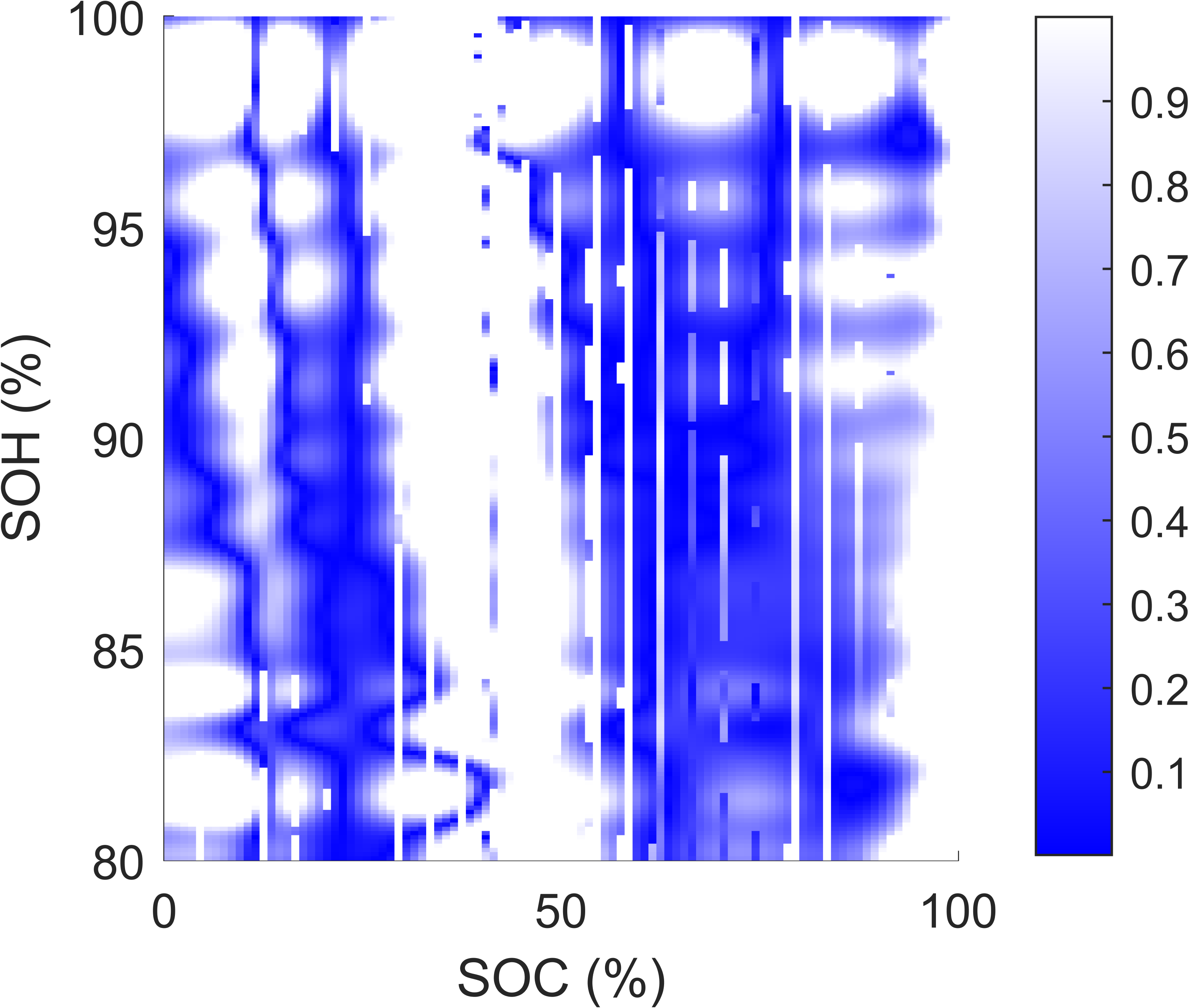}
\caption{Local convergence analysis}
\label{local}
\end{figure}

When $|L|<1$, we can conclude that the iteration is locally convergent, since $\frac{|SOH_{k+1}-SOH_{true}|}{|SOH_{k}-SOH_{true}|}<|L|+\epsilon<1$. On the other hand, when $L>1$, the method will not converge locally because in the neighborhood, $\frac{|SOH_{k+1}-SOH_{true}|}{|SOH_{k}-SOH_{true}|}>|L|-\epsilon>1$, and $SOH_k$ will finally move out of the neighborhood. The value of $|L|$ as a function of $SOC_{true}$ and $SOH_{true}$ is shown in Figure \ref{local}. In the figure, for better visualization, all the areas where $|L|>1$ are rescaled to $|L|=1$, and other areas represent the domain where the method is locally convergent.

\begin{figure}[htbp]
\centering
\includegraphics[width=9cm]{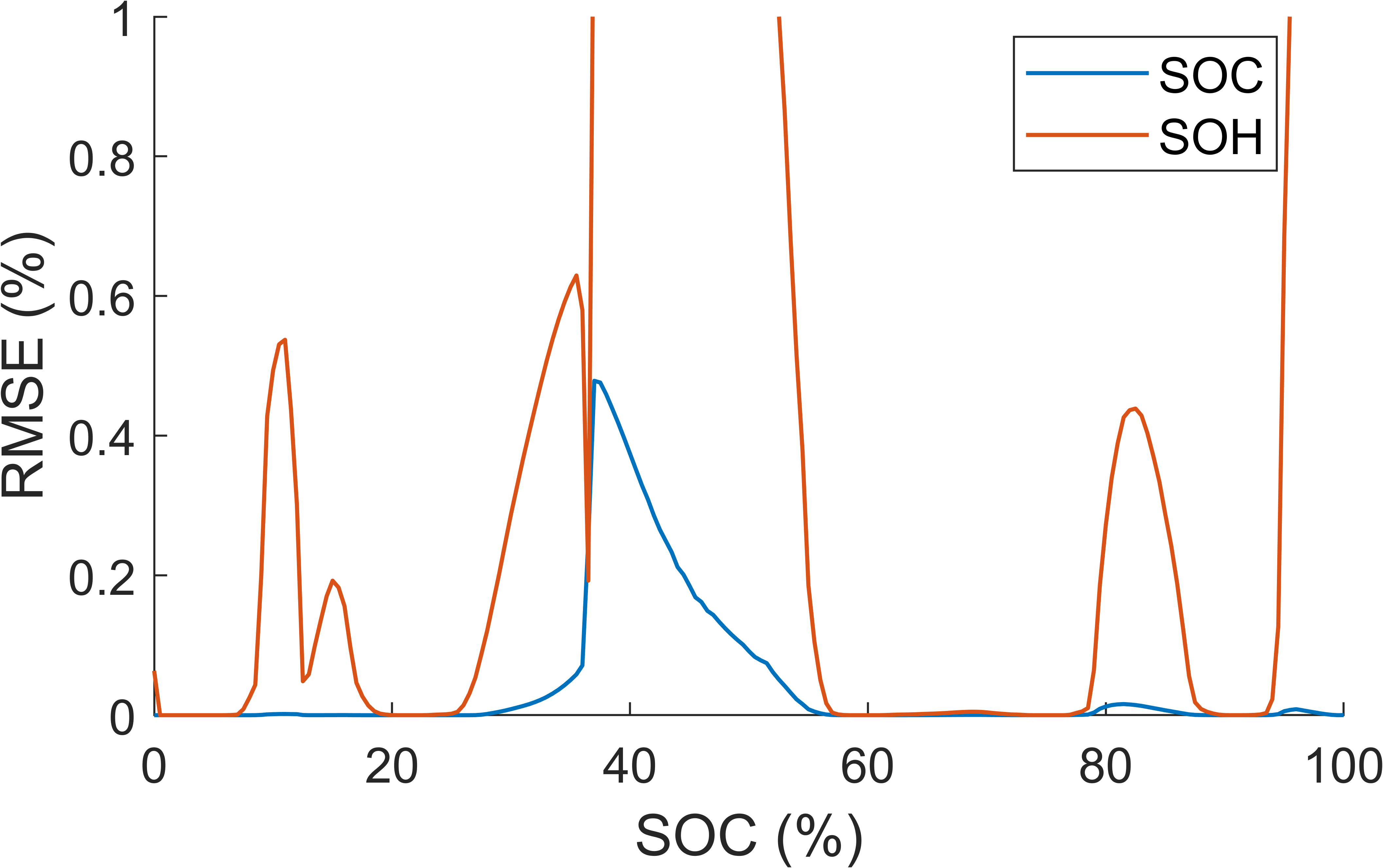}
\caption{Iteration error analysis}
\label{SOCSOHconvergence}
\end{figure}

The local convergence analysis can tell us what values of $SOC, SOH$ can be estimated, and what values cannot. For example, if the iterative method is used at 50\% $SOC$, it will almost always give inaccurate results since the iteration is not locally convergent at 50\% SOC, regardless of the true SOH. However, local convergence analysis alone is not enough to evaluate the estimation error since local convergence doesn't guarantee global convergence, and not converging at a certain point does not necessarily mean a high estimation error. To better evaluate the accuracy limit of the iterative method, another simulation was performed to analyze the relationship between estimation accuracy and $SOC_{true}$. Namely, for each $SOC_{true}$ between 0\% to 100\%, we enumerate the values of $SOH_{true}$ from 80\% to 100\% and calculate the average estimation error (measured by root mean square error, or RMSE) when the iterative method is initialized at $SOH_0=100\%$. The results are shown in Figure \ref{SOCSOHconvergence}.

According to Figure \ref{SOCSOHconvergence}, the estimation error of SOC is always relatively low, as the maximum RMSE of SOC is only 0.5\%. In contrast, the estimation error of SOH varies more significantly to $SOC_{true}$ when using the method. By comparing Figure \ref{local} with Figure \ref{SOCSOHconvergence}, we can see that the two figures are consistent with each other. Namely, Figure \ref{local} suggests that the algorithm is not locally convergent at around 50\% and 100\% SOC, regardless of the true SOH. In Figure \ref{SOCSOHconvergence}, we can see that the state estimation errors in these regions are indeed high. On the other hand, Figure \ref{local} suggests that the algorithm is usually locally convergent between 57\% to 77\% SOC, and Figure \ref{SOCSOHconvergence} tells us that global convergence is also usually guaranteed in this range. Since the error of SOC estimation can be higher in practice, the best timing to use our method is between 57\% and 77\% SOC, where the estimations can almost always converge to the correct values.
\section{Experimental validation}\label{section_val}
\subsection{Experimental validation setup}
\begin{figure}[ht]
\centering
\includegraphics[width=10cm]{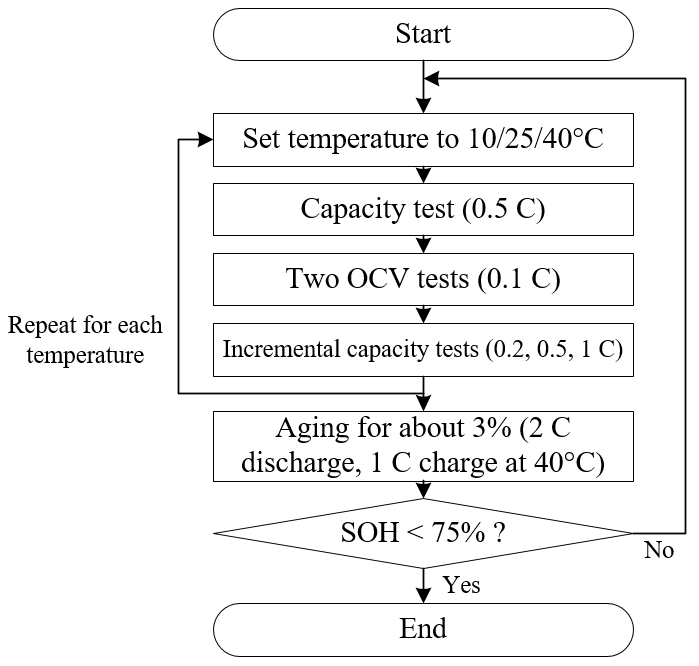}
\caption{A flow chart of the aging test}
\label{flow}
\end{figure}

Six lithium-ion batteries of NMC chemistry (model ICR18650-22F) manufactured by SAMSUNG were placed inside six separate temperature control chambers and were used for experimental validation. The charging and discharging were monitored by the battery tester manufactured by Arbin Instruments. Data collection and the setup of charging and discharging profiles were done using the supporting software MITS Pro. The RMSEs of the voltage and current measurements are 0.15 mV and 0.1 mA, respectively. The testing profile for each cell was precisely the same and is shown in Figure \ref{flow}.

As is shown in Figure \ref{flow}, the cells were aged at 40$^{\circ}$C by running continuous 1 C charge and 2 C discharge until the SOH dropped below 75\%. Characteristic tests were executed on each cell at nine (or more) different aging levels. Each characteristic test consisted of a capacity test (to stabilize the cell temperature), two OCV tests (incremental OCV test and low-current OCV test), and three incremental capacity tests at different C rates. All these characteristic tests were repeated at three different temperatures. The meanings of the two OCV tests have already been illustrated in Section \ref{OCV_section}. The charge and discharge rates used in the low-current OCV test and incremental OCV test were both 0.1 C. The relaxation time in the incremental OCV test was 3 minutes, much shorter than the typical setup. The purpose of shortening the relaxation time is to make the OCV in the OCV experiment more consistent with the OCV identified from the battery ECM and, hence, to improve the SOC and SOH estimation accuracy. In the incremental capacity tests, the cells were fully charged and discharged at a constant current until the SOC dropped by 3\%. Afterward, the cell rested for three minutes, followed by a 10-second 1 C discharge and another 3-minute rest. Then, the cell was discharged again. The procedure described above was repeated until the cell was fully discharged. The cell then rested for 30 minutes and was charged using a constant current constant voltage profile. Likewise, whenever the SOC rose by 5\%, the cell rested for 3 minutes, was discharged for 10 seconds, and rested for another 3 minutes until the cell was fully charged. The purpose of such a charging and discharging profile was to enable the validation of our parameter estimation algorithm at a suitable SOC where the algorithm can converge properly.

According to the conclusions in Section \ref{converge}, the best time to use the method is when the actual SOC is between 55\% and 77\%. However, in practice, the SOC is unknown before the state estimation, so the SOC is not a good indicator of when to use the method. Under this consideration, we instead used the terminal voltage as the indicator during validation. Namely, among more than twenty relaxations during the charging process in the incremental capacity test, the method is used when the terminal voltage at the beginning of the relaxation exceeds 3.9 V. This voltage threshold can guarantee that the SOC during this and the following relaxation are both between 55\% and 77\%, regardless of the SOH and the C rate. 

According to Proposition \ref{x3} and \ref{x2} in Section \ref{3points}, when estimating the ECM parameters, the optimal selection of the three points should be $x_1=0, x_2=x_3/2$, and 50s$\leq x_3 \leq$120s (depending on the estimation accuracy). However, Proposition \ref{x3} is based on the assumption that the ECM precisely describes the dynamic voltage response of the battery, which is, in fact, not true. In practice, it is found that the ECM model can only approximate the cell's voltage response between 10s to 300s well. As a result, we instead selected $x_1$=10s, $x_2=\frac{x_1+x_3}{2}$, and 60s$\leq x_3 \leq$180s. 

During the experimental validation, the OCV curve was fitted based on the voltage data in the incremental OCV test. Since the proposed SOC and SOH estimation method is used during the charging process, we only used the charge OCV data to fit the OCV curve, as the charge OCV curve is slightly different from the discharge OCV curve. After fitting the OCV curve, we validated our method on the experimental data in the three incremental capacity tests. Such a process was repeated at each temperature. As previously mentioned, when validating on a particular cell, the OCV data of this cell were excluded when fitting the OCV curve. For example, when estimating the SOC and SOH of Cell 3, the OCV curve we used was fitted to the data of Cells 1, 2, 4, 5, and 6. The reason for such an arrangement was to separate the fitting data and validation data.

\subsection{Experimental results}
\begin{figure}[htbp]
\centering
\includegraphics[width=9cm]{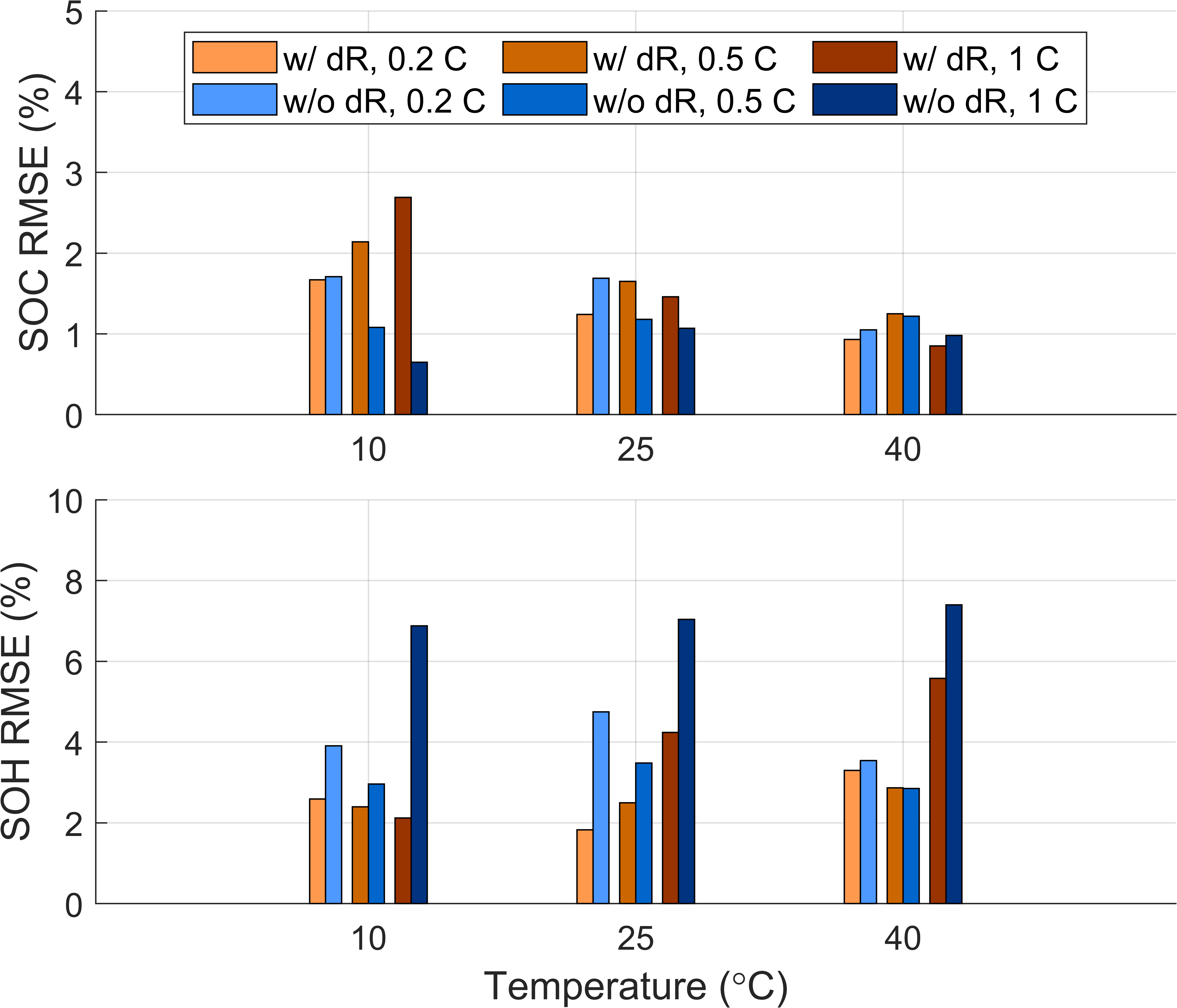}
\caption{SOC and SOH estimation errors of our method with and without $dR$ compensation (two-minute relaxation time)}
\label{result0}
\end{figure}

When $x_3=$120s, our method's SOC and SOH estimation results with and without $dR$ compensation are shown in Fig. \ref{result0}. The average run time of these two variants is, respectively, 0.33 ms and 0.51 ms.

%\begin{table}[htbp]
%\label{result0}
%\caption{SOC and SOH estimation errors of our method (without $dR$ compensation, two-minute relaxation time, average run time = 0.33 ms)}
%\centering
%\begin{tabular}{|c|c|c|c|}
%\hline
%SOC RMSE & 0.2 C & 0.5 C & 1 C \\ \hline
%10 $^{\circ}$C & 1.71\% & 1.08\% & 0.65\% \\ \hline
%25 $^{\circ}$C & 1.69\% & 1.18\% & 1.07\% \\ \hline
%40 $^{\circ}$C & 1.05\% & 1.22\% & 0.98\% \\ \hline
%SOH RMSE & 0.2 C & 0.5 C & 1 C \\ \hline
%10 $^{\circ}$C & 3.91\% & 2.96\% & 6.88\% \\ \hline
%25 $^{\circ}$C & 4.75\% & 3.48\% & 7.04\% \\ \hline
%40 $^{\circ}$C & 3.54\% & 2.85\% & 7.40\% \\ \hline
%\end{tabular}
%\end{table}

%\begin{table}[htbp]
%\label{result1}
%\caption{SOC and SOH estimation errors of our method (with $dR$ compensation, two-minute relaxation time, average run time = 0.51 ms)}
%\centering
%\begin{tabular}{|c|c|c|c|}
%\hline
%SOC RMSE & 0.2 C & 0.5 C & 1 C \\ \hline
%10 $^{\circ}$C & 1.67\% & 2.14\% & 2.69\% \\ \hline
%25 $^{\circ}$C & 1.24\% & 1.65\% & 1.46\% \\ \hline
%40 $^{\circ}$C & 0.93\% & 1.25\% & 0.85\% \\ \hline
%SOH RMSE & 0.2 C & 0.5 C & 1 C \\ \hline
%10 $^{\circ}$C & 2.59\% & 2.40\% & 2.12\% \\ \hline
%25 $^{\circ}$C & 1.83\% & 2.50\% & 4.24\% \\ \hline
%40 $^{\circ}$C & 3.30\% & 2.87\% & 5.58\% \\ \hline
%\end{tabular}
%\end{table}
According to Fig. \ref{result0}, the proposed method can realize fast and accurate SOC and SOH estimation under various temperatures and C rates. When the C rate is equal to or lower than 0.5 C, the variant without $dR$ compensation is enough for accurate estimation, with a SOC and SOH error of about 2\% and 3.5\%, respectively. Note that this variant only requires a relaxation of two minutes and a computational time of 0.33 ms. If the charging current is higher than 0.5 C or the requirement for estimation accuracy is higher, then using the variant with $dR$ compensation would be prudent. While the DR compensation variant is more complex, it still only needs a four-minute relaxation time and a computational time of 0.51 ms. This means that even if this variant is applied to an EV with 8,000 cells, the total computational time would still be lower than five seconds.

\begin{figure}[htbp]
\centering
\includegraphics[width=9cm]{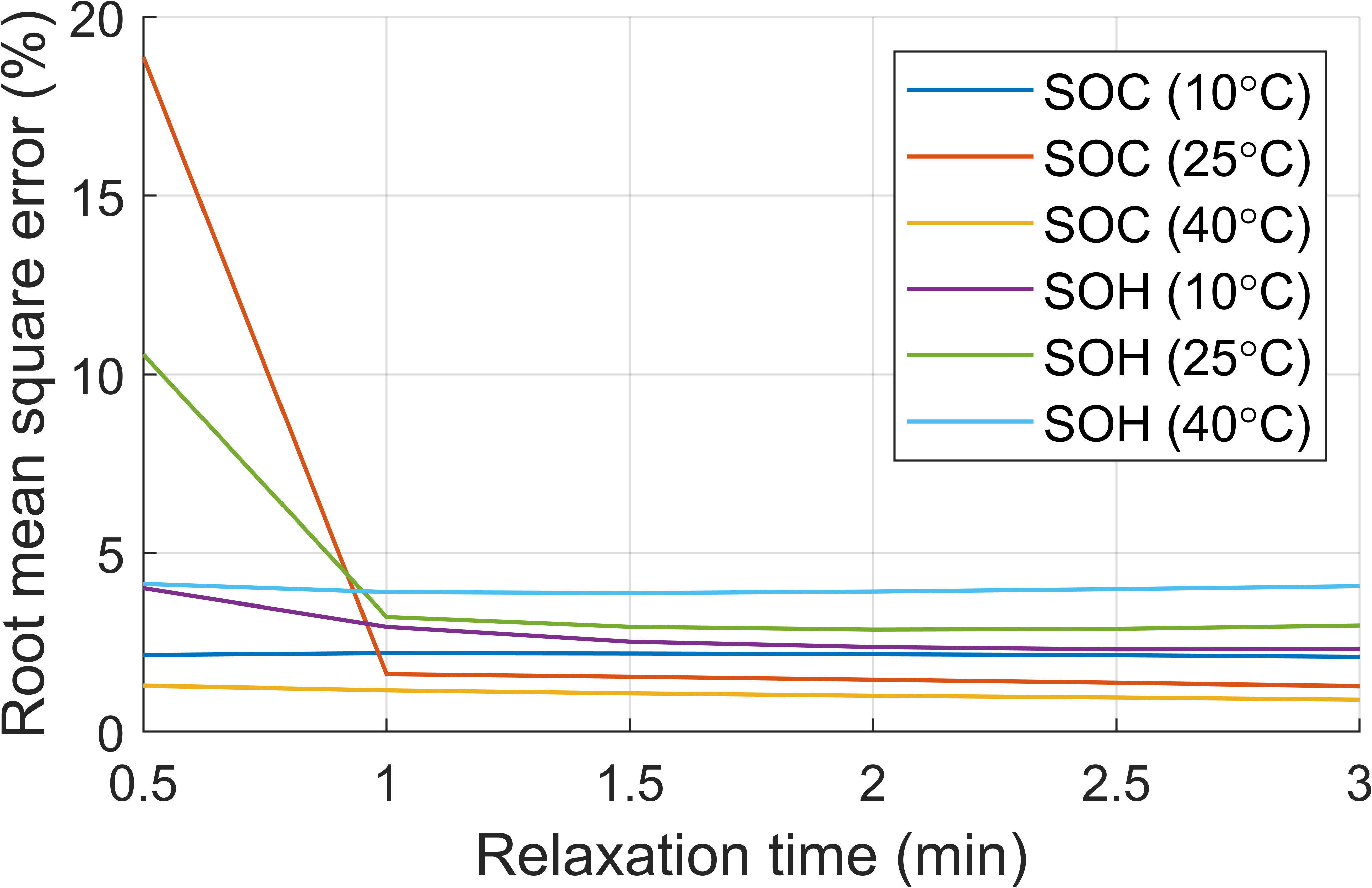}
\caption{Selection of the relaxation time ($x_3$)}
\label{x3_result}
\end{figure}

To verify the method's sensitivity to the relaxation time ($x_3$), the average SOC and SOH estimation errors for different relaxation times and temperatures are plotted in Fig. \ref{x3_result}. The results show that the estimation error is not very sensitive to the relaxation time after it exceeds 60 seconds. The lowest error is achieved at $x_3 = 120 $s, suggesting that relaxing the cells for more than two minutes is unnecessary.

\subsection{Comparison against Unscented Kalman Filter}
\label{comparison}
Compared with other model-based methods (for example, EKF and UKF), the proposed method has significantly less computational complexity and needs no hyperparameter tuning or initialization. However, it requires the cell to rest for one to four minutes during the charging process, while other model-based methods do not have such requirements. As a result, to better evaluate the pros and cons of the proposed method, it is necessary to make a quantitative comparison with other model-based methods in terms of their accuracy and computational complexity.

%In this paper, the UKF algorithm was chosen as the baseline algorithm. While another KF-based method was used in Section \ref{section_track} for SOC tracking, the purpose differed. In Section \ref{section_track}, EKF was only used to estimate the SOC of one cell for each battery string, under which circumstance all the states of the model are observable, and the convergence is guaranteed; but here, the UKF is used to estimate the SOC and SOH of each cell in the pack, and not the states are not always fully observable. 
In this paper, the UKF algorithm was chosen as the baseline algorithm. The UKF has three states: remaining capacity $Q_r$, capacitor voltage $U_c$, and present maximum capacity $Q$. With these definitions, the discrete state-space representation of the system can be written as (\ref{state_space}). The relationship between these states and the SOC and SOH is formulated in (\ref{SOC}) and (\ref{SOH}). Note that we avoid directly defining $SOC$ and $SOH$ as the states to make the state transition functions linear. 

\begin{equation}
\label{state_space}
\begin{cases}
    \begin{bmatrix} Q_{r,k}  \\ U_{c,k} \\ Q_{k}\end{bmatrix} = 
    F 
    \begin{bmatrix}Q_{r,k-1}  \\ U_{c,k-1} \\ Q_{k-1} \end{bmatrix} +B I_k\\
    U_{ter,k}=f_{OCV}\left(\frac{Q_{r,k}}{Q_{k}},\frac{Q_{k}}{Q_{0}}\right)+U_{c,k}+R_1I_k
\end{cases}    
\end{equation}
\begin{equation}\label{FB}
    F=\begin{bmatrix}1 & 0 & 0\\0 & e^{\frac{-\Delta t}{R_2C}} & 0\\0 &0 &1\end{bmatrix},B=\begin{bmatrix} \Delta t\\R_2-R_2e^{\frac{-\Delta t}{R_2C}}\\0 \end{bmatrix}
\end{equation}
where $\Delta t$ is the time interval between two steps, and $B$ and $F$ are two constant matrices defined by (\ref{FB}). 

The UKF was validated on the same data as our method (with $dR$ compensation) for a fair comparison. Specifically, the profile includes a three-minute relaxation, a 10-second 1 C discharge pulse, another 3-minute relaxation, a constant-current charging that increases the SOC by 5\%, and another 3-minute relaxation. During the first relaxation, the UKF was not turned on, and the voltage data was used to estimate the values of all the ECM parameters. The parameter estimation method is the same as in Sec. \ref{parameter_section}. Starting from the discharge pulse, the UKF was turned on, with the initial SOH estimation set to 100\%, the initial capacitor voltage set to 0 V, and the initial SOC set to be the inverse of the OCV during the first relaxation. The SOC and SOH estimation at the very end of the profile are considered estimation outputs and compared against the actual value to calculate the error. The detailed hyperparameter setup for this UKF is the same as in \cite{jiang2024}.

\begin{figure}[htbp]
\centering
\includegraphics[width=9cm]{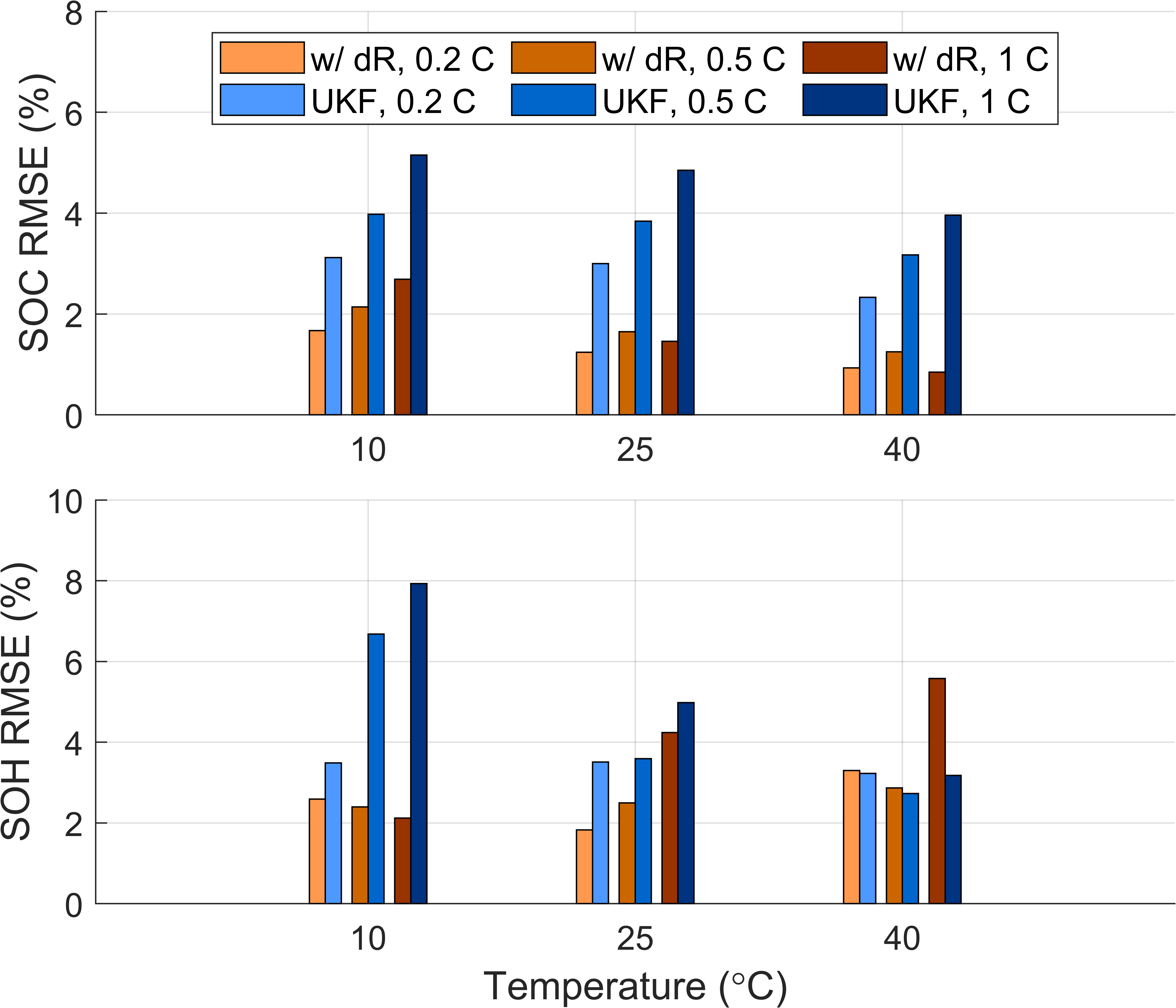}
\caption{Comparison between our method (average run time = 0.51 ms) and UKF (average run time = 73 ms)}
\label{resultUKF}
\end{figure}

The SOC and SOH estimation results from the UKF are presented in Fig. \ref{resultUKF}. The figure suggests that our method is hundreds of times faster than the UKF and is more accurate, especially for SOC estimation.

%\begin{table}[htbp]
%\label{resultUKF}
%\caption{SOC and SOH estimation errors of UKF and their ratio to our method (average run time = 73 ms)}
%\centering
%\begin{tabular}{|c|c|c|c|}
%\hline
%SOC RMSE & 0.2 C & 0.5 C & 1 C \\ \hline
%10 $^{\circ}$C & 3.12\% ($\times$1.9)& 3.98\% ($\times$1.9)& 5.15\% ($\times$1.9)\\ \hline
%25 $^{\circ}$C & 3.00\% ($\times$2.4)& 3.84\% ($\times$2.3)& 4.85\% %($\times$3.3)\\ \hline
%40 $^{\circ}$C & 2.33\% ($\times$2.5)& 3.17\% ($\times$2.5)& 3.96\% ($\times$4.7)\\ \hline
%SOH RMSE & 0.2 C & 0.5 C & 1 C \\ \hline
%10 $^{\circ}$C & 3.49\% ($\times$1.3)& 6.68\% ($\times$2.8)& 7.93\% ($\times$2.9)\\ \hline
%25 $^{\circ}$C & 3.51\% ($\times$1.9)& 3.59\% ($\times$1.4)& 4.98\% ($\times$1.2)\\ \hline
%40 $^{\circ}$C & 3.23\% ($\times$1.0)& 2.73\% ($\times$1.0)& 3.18\% ($\times$0.6)\\ \hline
%\end{tabular}
%\end{table}
\subsection{Error analysis}
While the SOC and SOH estimation error of the proposed method is relatively low, it is still essential to understand the source of the error. Understanding the error sources can help us better analyze the applicability and the pros and cons of the method. For this purpose, the validation was done another eighteen times, each time on a different setup. As shown in the first column of Table \ref{error}, these extra validations can be separated into three groups according to the data source. In the first group, all the validations were done on experimental data. In the second and third groups, only the current data were from the experiment, and all the voltage data came from a simulated battery model. The difference is that the battery model used in the second group is a second-order RC model, while the one used in the third group is a first-order RC model. In other words, only the model used in the third group is the same as the battery model we used when proposing the method. By default, when generating the voltage data, it is assumed that there is a voltage noise with a standard deviation of 0.15 mV (the same as the standard deviation of the voltmeter in the experiment), and all the RC parameters change as the SOC changes.

We started with the default setting for each data source and gradually added more simplifications. For example, in the fourth scenario, we not only used the simplification ``Known SOC'' but also applied the simplification ``More relevant data'' and ``Fixed OCV curve''. In other words, within each group in Table \ref{error}, the difficulty of the estimation becomes gradually less from top to bottom. Specifically, for experimental data, ``More relevant data'' means that the OCV curve we used came from the same cell that we validating on (note that by default, the OCV curve came from the data of all the other cells besides the cell that we were validating on). ``Fixed OCV curve'' means that we stopped updating the OCV curve as we updated the SOH estimation, and always used the OCV curve corresponding to the true SOH. ``Known SOC'' means that we assume that we used the actual SOC instead of the estimated SOC when estimating SOH. For simulation data, ``Known capacitor voltage'' means that the actual value of the capacitor voltage before entering the relaxation, which was previously assumed to be $R_2 I$, is known. ``No voltage noise'' means that the measurement voltage noise is set to zero. ``Fixed RC parameter'' means that the RC parameters in the simulation become constants instead of SOC-dependent variables. ``More accurate dV/dt'' means that fewer points (two points instead of fifty points) are used in the linear regression to fit dV/dt. The meaning of ``Fixed OCV curve'' and ``Known SOC'' for simulation data is the same as for the experimental data.

The SOC and SOH estimation errors for each validation case are presented in Table \ref{error}. Note that all the validations were based on data gathered at 25 $^{\circ}$C, and we always added $dR$ compensation.
\begin{table}[htbp]\scriptsize
\label{error}
\caption{RMSE of SOC and SOH estimation under different scenarios}
\centering
\begin{tabular}{|c|c|cc|cc|cc|}
\hline
\multirow{2}{*}{Data Source} & \multirow{2}{*}{Scenarios} & \multicolumn{2}{c|}{0.2 C} & \multicolumn{2}{c|}{0.5 C} & \multicolumn{2}{c|}{1 C} \\ \cline{3-8} 
 &  & \multicolumn{1}{c|}{SOC} & SOH & \multicolumn{1}{c|}{SOC} & SOH & \multicolumn{1}{c|}{SOC} & SOH \\ \hline
\multirow{5}{*}{\begin{tabular}[c]{@{}c@{}}Experimental\\ Data\end{tabular}} & Default & \multicolumn{1}{c|}{1.24\%} & 1.83\% & \multicolumn{1}{c|}{1.65\%} & 2.50\% & \multicolumn{1}{c|}{1.46\%} & 4.24\% \\ \cline{2-8} 
 & $\cdots$ + More relevant data & \multicolumn{1}{c|}{1.24\%} & 1.84\% & \multicolumn{1}{c|}{1.69\%} & 2.55\% & \multicolumn{1}{c|}{1.57\%} & 4.23\% \\ \cline{2-8} 
 & $\cdots$ + Fixed OCV curve & \multicolumn{1}{c|}{0.62\%} & 2.31\% & \multicolumn{1}{c|}{1.35\%} & 2.84\% & \multicolumn{1}{c|}{1.88\%} & 4.44\% \\ \cline{2-8} 
 & $\cdots$ + Known SOC & \multicolumn{1}{c|}{0.62\%} & 2.33\% & \multicolumn{1}{c|}{1.35\%} & 2.79\% & \multicolumn{1}{c|}{1.88\%} & 4.85\% \\ \cline{2-8} 
 & Coulomb Counting & \multicolumn{1}{c|}{0.20\%} & 0.68\% & \multicolumn{1}{c|}{0.77\%} & 1.79\% & \multicolumn{1}{c|}{0.38\%} & 2.06\% \\ \hline
\multirow{7}{*}{\begin{tabular}[c]{@{}c@{}}2RC Model \\ Simulation\end{tabular}} & Default & \multicolumn{1}{c|}{0.78\%} & 2.92\% & \multicolumn{1}{c|}{0.89\%} & 7.50\% & \multicolumn{1}{c|}{1.40\%} & 11.72\% \\ \cline{2-8} 
 & $\cdots$ + Known capacitor voltage & \multicolumn{1}{c|}{1.29\%} & 2.27\% & \multicolumn{1}{c|}{2.37\%} & 4.18\% & \multicolumn{1}{c|}{3.70\%} & 6.89\% \\ \cline{2-8} 
 & $\cdots$ + No voltage noise & \multicolumn{1}{c|}{1.14\%} & 1.71\% & \multicolumn{1}{c|}{2.55\%} & 3.88\% & \multicolumn{1}{c|}{3.74\%} & 6.94\% \\ \cline{2-8} 
 & $\cdots$ + Fixed RC parameters & \multicolumn{1}{c|}{0.35\%} & 0.52\% & \multicolumn{1}{c|}{1.15\%} & 0.81\% & \multicolumn{1}{c|}{3.14\%} & 3.39\% \\ \cline{2-8} 
 & $\cdots$ + More accurate dV/dt & \multicolumn{1}{c|}{0.35\%} & 0.52\% & \multicolumn{1}{c|}{1.14\%} & 0.77\% & \multicolumn{1}{c|}{3.12\%} & 3.32\% \\ \cline{2-8} 
 & $\cdots$ + Fixed OCV curve & \multicolumn{1}{c|}{0.31\%} & 0.14\% & \multicolumn{1}{c|}{0.93\%} & 0.42\% & \multicolumn{1}{c|}{2.22\%} & 2.39\% \\ \cline{2-8} 
 & $\cdots$ + Known SOC & \multicolumn{1}{c|}{0.31\%} & 0.08\% & \multicolumn{1}{c|}{0.93\%} & 0.09\% & \multicolumn{1}{c|}{2.22\%} & 0.20\% \\ \hline
\multirow{7}{*}{\begin{tabular}[c]{@{}c@{}}1RC Model \\ Simulation\end{tabular}} & Default & \multicolumn{1}{c|}{0.72\%} & 2.57\% & \multicolumn{1}{c|}{1.24\%} & 6.11\% & \multicolumn{1}{c|}{2.12\%} & 11.41\% \\ \cline{2-8} 
 & $\cdots$ + Known Capacitor Voltage & \multicolumn{1}{c|}{1.05\%} & 2.40\% & \multicolumn{1}{c|}{1.44\%} & 2.44\% & \multicolumn{1}{c|}{1.26\%} & 4.28\% \\ \cline{2-8} 
 & $\cdots$ + No voltage noise & \multicolumn{1}{c|}{0.85\%} & 1.13\% & \multicolumn{1}{c|}{1.40\%} & 1.80\% & \multicolumn{1}{c|}{1.20\%} & 3.93\% \\ \cline{2-8} 
 & $\cdots$ + Fixed RC parameters & \multicolumn{1}{c|}{0.10\%} & 0.50\% & \multicolumn{1}{c|}{0.16\%} & 0.62\% & \multicolumn{1}{c|}{0.53\%} & 1.25\% \\ \cline{2-8} 
 & $\cdots$ + More accurate dV/dt & \multicolumn{1}{c|}{0.11\%} & 0.51\% & \multicolumn{1}{c|}{0.17\%} & 0.62\% & \multicolumn{1}{c|}{0.47\%} & 1.10\% \\ \cline{2-8} 
 & $\cdots$ + Fixed OCV curve & \multicolumn{1}{c|}{0.00\%} & 0.00\% & \multicolumn{1}{c|}{0.00\%} & 0.00\% & \multicolumn{1}{c|}{0.00\%} & 0.00\% \\ \cline{2-8} 
 & $\cdots$ + Known SOC & \multicolumn{1}{c|}{0.00\%} & 0.00\% & \multicolumn{1}{c|}{0.00\%} & 0.00\% & \multicolumn{1}{c|}{0.00\%} & 0.00\% \\ \hline
\end{tabular}
\end{table}

Interestingly, when using our method on experimental data, none of the simplifications result in improved SOH accuracy. Only the ``Fixed OCV curve'' marginally improves SOC accuracy. This phenomenon suggests that all the error sources we examined in the first validation group are unimportant. According to the results in the second and third groups of validations, the critical error sources of our method are: (1) assuming the capacitor has been fully charged before the start of the relaxation and (2) the linear approximation of $R_1(SOC)$ and $R_2(SOC)$. The second error source is difficult to avoid since estimating $R_1$ and $R_2$ in real-time is challenging. Yet, the first error source can be mitigated by prolonging the constant-current charging time before the relaxation. In the experiment, the length of constant-current charging is set to the time required to increase the SOC by 5\%. Therefore, when the charge C rate is 1 C, the constant-current charging time is just 3 minutes, which is insufficient for the capacitor to charge fully. Such a difference in the length of constant-current charging explains why this error source affects higher C rates more than lower C rates. The estimation accuracy is expected to increase if we prolong the constant-current charging time. 

Additionally, by comparing the scenarios in the second and third data sources, we see that the estimation accuracy between these two data sources is, in fact, quite similar. This holds true even though the first-order RC model used in the parameter estimation only matches the third data source well. This observation suggests that although a more complex battery model can more accurately predict voltage data, the simple first-order RC model will not necessarily lead to a greater SOC and SOH estimation error. Other error sources can dilute the benefit of a more complex model, and identifying more parameters also leads to weaker observability and longer computational time.

Among all the scenarios in Table \ref{error}, the most unique one is ``Coulomb Counting''. For this setup, we did not use our SOC and SOH estimation method; instead, we used Coulomb Counting. As previously mentioned, the ``true'' SOC and ``true'' SOH are also calculated based on Coulomb Counting in the experimental validation. The difference is that our definition of SOH is the cell's normalized charge capacity in the 0.1 C incremental OCV test. Yet, here, the SOH was estimated by the normalized charge capacity in the corresponding incremental capacity test, in which the charge current is not 0.1 C. For example, when the charge current is 0.1 C, the initial capacity of Cell 1 is 2.156 Ah, which decreases to 1.775 Ah after 900 cycles, resulting in an SOH of 82.3\%. Meanwhile, when the maximum charge current is 1 C, the initial capacity of Cell 1 is 2.079 Ah, which decreases to 1.627 Ah after 900 cycles, resulting in an SOH of 78.3\%. This 4\% difference illustrates how SOH is not a universally well-defined quantity and depends on the specific definition. The estimated SOH RMSE is used to quantify the uncertainty originating from the definition of SOH itself. Likewise, when we calculate the SOC based on the charge and discharge capacity calculated from the 0.1 C OCV test, the SOC at the end of charging may not be precisely 100\%, since the cell's capacity varies slightly at different C rates. The RMSE of the SOC estimation is calculated from the difference between the final calculated SOC and 100\%, and is used to quantify the uncertainty originating from the definition of SOC. In essence, the error of Coulomb Counting can be considered the lower limit of the SOC and SOH estimation error when using experimental data since such an error comes from the definition of SOC and SOH and is impossible to avoid.

\section{Conclusions}
In this paper, an online battery SOC and SOH estimation method was proposed, which requires no hyperparameter tuning. Various theoretical analyses elucidate the algorithm's properties. The method has two variants. The first variant is designed for low C-rate ($\leq$0.5 C) charging and only requires a 1 to 2-minute relaxation. The second variant can be used for higher C-rates, but requires a 2 to 4-minute relaxation during the charging process. Both variants can accurately estimate SOC and SOH at different temperatures. Specifically, the average computational time of the first variant is 0.33 ms, and the estimation RMSE for SOC and SOH are respectively around 1.2\% and 4.8\%. In comparison, the average computational time of the second variant is 0.51 ms, and the estimation RMSE for SOC and SOH are respectively around 1.5\% and 3\%. Compared with UKF, our method requires significantly lower computational time and has higher accuracy. 

The main limitations of the work are that the proposed method requires a few minutes of relaxation at some specific SOC range (between 55\% and 77\%). It also requires low to moderate charging currents ($<1$ C) for accurate state estimation. Future work will focus on extending the method to a broader SOC range and fast-charging scenarios.

\section{CRediT authorship contribution statement}
\textbf{Shida Jiang:} Conceptualization, Data curation, Investigation, Formal analysis, Visualization, Methodology, Project administration, Writing – original draft, Writing – review \& editing. \textbf{Junzhe Shi:} Writing – original draft, Writing – review \& editing. \textbf{Scott Moura:} Resources, Funding acquisition, Project administration, Supervision, Writing – review \& editing.
\section{Declaration of competing interest}
The authors declare that they have no known competing financial interests or personal relationships that could have appeared to influence the work reported in this paper. 
%\section{Acknowledgment}
%This research did not receive any specific grant from funding agencies in the public, commercial, or not-for-profit sectors.
\section{Data availability}
The data and codes are available at this \href{https://berkeley.box.com/s/jz1w6po2iqzzfy7irxd9ok47ku3tr86j}{url}.
\appendix
\section{An example of SOC tracking method}\label{section_track}
As shown in Figure \ref{algorithm}, the proposed SOC and SOH estimation method outputs the SOC and SOH during the first relaxation period. The method can be repeated after several cycles to keep track of the SOH. As for the fast-changing SOC, however, an additional method is necessary to keep tracking its value after the first estimation is made during the relaxation. Noticing that the current is the same for all the cells connected in series in a battery pack. Therefore, according to (\ref{dSOC}), the SOC of different cells has the following relationship:
\begin{equation}
\label{pack_SOC}
    \Delta SOC_j=\Delta SOC_i\times \frac{SOH_i}{SOH_j}
\end{equation}
where $\Delta SOC_i$ is the change of SOC of Cell $i$ after the SOC and SOH estimation is done. According to (\ref{pack_SOC}), the SOC of each cell in the string can be calculated once the SOC of any single cell is estimated and the SOH of other cells is known. Since the SOC tracking method here does not need to be repeated for every cell, it can be a complex, filter-based method. 

In algorithm \ref{tracking}, an example of using an EKF for SOC tracking is provided. The method selects SOC and the capacitor voltage of Cell $i$ (the reference cell) as the states of the system. Note that all the inputs of the algorithms come from the measurement data and the output of the SOC and SOH estimation algorithm, as presented in Figure \ref{algorithm}.

\begin{algorithm}
\caption{SOC tracking algorithm based on extended Kalman filter}\label{tracking}
\textbf{Inputs}: $\bm{SOC_0}, \bm{SOH}, R_1, R_2, C, I_k, U_{ter,k},Temp_k,t_k$
%\textbf{Hyperpameters}: $P_{0},R_k,Q_k$ 
\begin{algorithmic}
\State $k = 0$ 
\While{the BMS is on}
\State $k = k+1$
\State $\Delta t = t_{k}-t_{k-1}$
\State  $\begin{bmatrix}SOC_{k}(i) \\U_{c,k} \end{bmatrix}  = \begin{bmatrix}1 & 0 \\0 & e^{-\Delta t/(R_2C)} \end{bmatrix}
\begin{bmatrix}SOC_{k-1}(i) \\U_{c,k-1} \end{bmatrix}$ + 
$\begin{bmatrix} \frac{\Delta t}{Q_{0}SOH_i} \\R_2(1-e^{-\Delta t/(R_2C)}) \end{bmatrix}I_k$
\State $P_k= \begin{bmatrix}1 & 0 \\0 & e^{-2\Delta t/(R_2C)} \end{bmatrix}P_{k-1}+Q_k$
\State $y_k = U_{ter,k}-OCV(SOC_k(i),SOH_i,Temp_k)-U_{c,k}-R_1I_k$
\State $H_k = \begin{bmatrix} \left.\dfrac{\partial OCV}{\partial SOC}\right|_{SOC=SOC_k(i),Temp=Temp_k} & 1\end{bmatrix}$
\State $S_k = H_kP_kH_k^T+R_k$
\State $K_k = P_kH_k^TS_k^{-1}$
\State $\begin{bmatrix}SOC_{k}(i) \\U_{c,k} \end{bmatrix}= \begin{bmatrix}SOC_{k}(i) \\U_{c,k} \end{bmatrix} +K_ky_k$
\State $P_k= (I-K_kH_k)P_k$
\For{$j =$ 1 to $N$}
        \State $SOC_{k}(j)=SOC_{0}(j)+(SOC_{i,k}-SOC_{0}(i))\times \frac{SOH_i}{SOH_j}$
\EndFor
\State \textbf{Output} $\bm{SOC_k}$
\EndWhile
\Statex where $N$ is the total number of cells in the battery pack. $i$ can be chosen arbitrarily. The algorithm inputs include the measured temperature, voltage, current, estimated SOC (during the relaxation), and estimated SOH of each cell. For Cell $i$ (the reference cell), its estimated ECM parameters are also required as the inputs. The output is the SOC of each cell in the battery pack. $R_k$ and $Q_k$ are, respectively, the process noise covariance matrix and the measurement covariance matrix, and they can be determined based on the method presented in \cite{jiang2024}.
\end{algorithmic}
\end{algorithm}

It is worth noting that, in Algorithm \ref{tracking}, all the parameters in the ECM are considered to be constants, which is actually not true in reality. Nevertheless, the algorithm can still yield a satisfactory result if the possible parameter changes are considered part of the measurement and process noise. Additionally, it is easy to prove that when the sampling frequency is high enough (in which case the linearization of the output function in each step can be considered perfect), all the states in the model are observable, and the convergence is therefore guaranteed. By contrast, suppose that a model parameter (such as the internal resistance) is instead considered to be a time-varying third state. In that case, the system's observability would depend on the input, and there may be no guarantee for convergence even if the sampling frequency is high enough. Therefore, only two states are included in the EKF.


\bibliographystyle{elsarticle-num} 
\bibliography{elsarticle-template-num}


%\begin{thebibliography}{00}


%\bibitem{}

%\end{thebibliography}

%\section{Confidence interval of the SOC and SOH estimation}

\end{document}